\documentclass[fleqn,usenatbib]{mnras}

\usepackage{newtxtext,newtxmath}

\usepackage[T1]{fontenc}

\DeclareRobustCommand{\VAN}[3]{#2}
\let\VANthebibliography\thebibliography
\def\thebibliography{\DeclareRobustCommand{\VAN}[3]{##3}\VANthebibliography}

\usepackage{amsmath}
\usepackage[nopatch]{microtype}
\usepackage{booktabs}
\usepackage{graphicx}
\usepackage{subcaption}
\usepackage{makecell}
\usepackage{hyperref}
\usepackage{pdflscape}
\usepackage{siunitx}
\newcommand{\XY}[2]{\left[\textrm{#1/#2}\right]}
\newcommand{\FeH}{\XY{Fe}{H}}
\newcommand{\XFe}[1]{\XY{#1}{Fe}}
\newcommand{\XH}[1]{\XY{#1}{H}}

\newcommand{\Teff}{T_{\textrm{eff}}}
\newcommand{\logg}{\log g}

\newcommand{\vmic}{v_{\rm mic}}
\newcommand{\vsini}{v \sin i}

\newcommand{\aFe}{\rm [\alpha/Fe]}

\defcitealias{lowe_detailed_2026}{Paper I}

\title[Metal-Poor Stars with X-Shooter II.]{Detailed Abundance Determination of Metal-Poor Stars with X-Shooter II. - Chemically Disentangling the Halo, Disk and GSE}

\author[B. Lowe et al.]{Benjamin D. C. Lowe,$^{1}$\thanks{E-mail: ben.lowe@anu.edu.au}
Luca Casagrande,$^{1}$
Thomas Nordlander,$^{2}$ 
Gary S. Da Costa$^{1}$
and Norbert Christlieb$^{3}$
\\
$^{1}$Research School of Astronomy and Astrophysics, Australian National University, Canberra, ACT 2611, Australia\\
$^{2}$Theoretical Astrophysics, Department of Physics and Astronomy, Uppsala University, Box 516, 751 20 Uppsala, Sweden\\
$^{3}$Zentrum für Astronomie der Universität Heidelberg, Landessternwarte, Königstuhl 12, 69117 Heidelberg, Germany\\
}

\date{Accepted XXX. Received YYY; in original form ZZZ}

\pubyear{\the\year{}}

\begin{document}
\label{firstpage}
\pagerange{\pageref{firstpage}--\pageref{lastpage}}
\maketitle

\begin{abstract}
We present a detailed chemical analysis of seven extremely metal-poor (EMP) star candidates observed with X-Shooter, combining them with $16$ EMP candidates from Paper I. We measured abundances for $16$ elements, showing excellent agreement with previously published results. The sample was further extended using high-resolution literature data for 315 metal-poor stars. The full sample was then kinematically separated into prograde disk, retrograde disk, Gaia-Sausage Enceladus (GSE) and halo classifications. Combining dynamics with chemistry, we demonstrate that the prograde disk exhibits a distinct negative linear trend in $\XY{Sc}{Mg}$ with increasing metallicity, with a slope of $-0.6$\,dex per dex. This contrasts with the halo trend at $-0.04$\,dex per dex, a difference significant at the $3.95\sigma$ level. Within the metallicity range $-4.2 \leq [\mathrm{Fe}/\mathrm{H}] \leq -1.9$, the prograde disk trend is driven by low $\XFe{Mg}$ at lower metallicities, along with low $\XFe{Sc}$ at higher metallicities. This could be due to reduced early Mg enrichment in the progenitor prograde disk, followed by subsequent Mg enrichment, possibly associated with a later gas accretion event. However, the physical origin of the higher-metallicity Sc depletion remains unexplained by current nucleosynthesis models. The result remains significant at the $>3\sigma$ level across kinematic classifications derived from a different Galactic potential. Additionally, we also identified an r-I star with enhanced Ti, moderately-enhanced Sc, and depleted in C (unrelated to its evolutionary state). These abundances suggest a massive jet-induced hypernova progenitor, though a measurement of Zn is needed to verify this.
\end{abstract}




\section{Introduction}
\label{sec: introduction}
Extremely metal-poor (EMP) stars, defined as having $\FeH < -3.0$\footnote{$[\textrm{X}/\textrm{H}] = \textrm{log}(N_{\textrm{X}}/N_{\textrm{H}})_{\star} - \textrm{log}(N_{\textrm{X}}/N_{\textrm{H}})_{\odot}$, where $N_{\textrm{X}}$ is the number density for element X.} are some of the oldest stellar objects we can observe today in the Universe \citep{beers_discovery_2005}, with some EMP stars enriched exclusively by supernovae from the very first stars, the metal-free Population III (Pop. III) stars \citep[e.g.][]{klessen_first_2023}. The atmospheres of these EMP stars preserve the state of their environment during their formation \citep[e.g.][]{caffau_x-shooter_2011, caffau_primordial_2012, sestito_exploring_2021, dovgal_probing_2024}, so by studying their atmospheres and kinematics, we can provide vital observational constraints on the evolution and formation of the different kinematic regions in our Galaxy at early times. Recent work has shown the existence of a ``primordial'' disk in their samples of very metal-poor stars (VMP; $-3.0 < \FeH \leq -2.0$), a separate component that was not formed solely from minor mergers or Galactic building blocks \citep[e.g.][]{sestito_tracing_2019, cordoni_exploring_2021, xu_additional_2025}. This ``primordial'' disk has also been suggested at lower metallicities by \citet{hong_candidate_2024}, who combined stars from the SkyMapper Southern Survey \citep[SMSS;][]{onken_skymapper_2019} and the Stellar Abundance and Galactic Evolution Survey \citep[SAGES;][]{fan_stellar_2023} to show that between $12.4$\,\% and $18.3$\,\% of all EMP stars possess highly prograde disk orbits. The existence of a rotation-supported disk with high $V_\phi$ (seen in the Milky Way today) has been shown to emerge around $\FeH \sim -1.3$ \citep{zhang_existence_2024}. This is supported with the spin-up of the Galactic disk being observed across metallicities $-1.5$ and $-1.1$\,dex \citep{lowe_detailed_2026}.


However, to understand the origin of the early Galactic disk population, work from high resolution cosmological zoom-in simulations is needed to compliment the observational analysis. Such work includes \citet{sestito_exploring_2021} \citep[using NIHAO-UHD;][]{buck_nihao-uhd_2020} and \citet{santistevan_origin_2021} \citep[using FIRE-2;][]{hopkins_fire-2_2018}, who both showed that in the earliest stages, the Galactic assembly was comprised of `building blocks' that merged together. Here, the depth of the gravitational potential well of the merging proto-Galaxy was low, dispersing the early stars into different kinematic configurations. At later times, the proto-Galaxy grew, forming the Galactic disk. Accreted systems at these times primarily contributed to the halo, but these simulations also showed that some of these systems deposited stars into the plane by dynamical friction. However, it has been shown by \citet{sotillo-ramos_likelihoods_2023} \citep[using TNG50;][]{nelson_first_2019, pillepich_first_2019} that \textit{in situ} formation is also likely for the metal-poor prograde disk population. These conflicting scenarios between accretion and \textit{in situ} formation of the early disk population mandate the need for detailed chemical abundances of metal-poor prograde disk stars, to break this degeneracy and uncover the true origin of the Milky Way's oldest planar structures.

This was attempted recently by \citet{sestito_ancient_2026}, who showed that their sample of very metal-poor (VMP; $\FeH < -2.0$) prograde and retrograde disk stars had $\XFe{Sr, Ba, Eu}$ abundances consistent with dwarf galaxies, hinting at possible accretion from a single system early in the history of the Galaxy. Though, because this study only looked at VMP stars and not the more metal-poor EMP populations, the earliest epochs of the disk's formation remains largely unstudied. To help with this, in \citet{lowe_rise_2025} we started a survey of disk metal-poor stars in the Galaxy using the 2dF multi-fibre positioner coupled with the low-resolution AAOmega spectrograph on the Anglo-Australian Telescope (AAT) \citep{saunders_aaomega_2004, sharp_performance_2006}. The most metal-poor candidates showed a substantial degree of chemical diversity and were subsequently observed with the medium-resolution X-Shooter spectrograph on the Very Large Telescope (VLT) \citep{vernet_x-shooter_2011} in \citet[][Paper I]{lowe_detailed_2026}. Here we present X-Shooter results for an additional $7$ EMP candidate stars. We then combine the full sample of $23$ X-shooters stars with abundance results from the literature for an additional $315$ metal-poor stars, studied at high-resolution, to enable the interpretation of our results in a broader context.

In what follows we present our X-Shooter observations, reduction and data analysis techniques (Section~\ref{sec: xshooter data}), then introduce the high-resolution literature sample (Section~\ref{sec:literature comparison}). We then present our X-Shooter chemical abundances (Section~\ref{sec:xshooter results}), followed by the chemodynamical results (Section~\ref{sec:chemodynamic results}). Finally, we discuss the key findings (Section~\ref{sec:discussion}).

\section{Observational Data}
\label{sec: xshooter data}
\subsection{Target Selection, observations and data reduction}
\label{subsec: target selection}
The seven metal-poor stars in this work (see Table~\ref{tab:star properties}) were observed as part of program ID 115.28C7.001 (hereafter P115 sample) from April to September 2025 on the high-efficiency X-Shooter spectrograph on Unit Telescope 2 (UT2, Kueyen) of the Very Large Telescope (VLT) at Cerro Paranal Observatory. Both Ultraviolet-Blue (UVB; $3000$--$5595$\,\AA{}) and the Visible (VIS; $5595$--$10240$\,\AA{}) arms were used, adopting slit widths of $1.0$\,arcsec for the UVB and VIS arms, respectively, and a slit length of $11$\,arcsec in both cases. These yielded resolving powers of $R = 5400$ and $8900$, respectively. These P115 stars accompany the $16$ stars from \citetalias[][]{lowe_detailed_2026}, observed on the same instrumental setup from April to July 2024 as part of program 113.26N5.001 (hereafter P113 sample). 

\begin{table*}
    \centering
    \caption{List of P115 stars studied. Their \textit{Gaia} DR3 source ID, RA and Dec coordinates (in the ICRS reference frame at the J2016.0 epoch), galactic coordinates (galactic longitude, $l$, and galactic latitude, $b$), parallax $\pi$, apparent magnitudes $\rm G$, reddening $E(B-V)$ (from \citet{schlegel_maps_1998}, rescaled as per \citet{casagrande_skymapper_2019}) and kinematic groupings from \citet{lowe_rise_2025} are provided (see Section~\ref{subsubsec: orbital analysis}).}
    \setlength{\tabcolsep}{3.8pt}
    \begin{tabular}{ll|crrccccc}
        \hline
        Star ID & \textit{Gaia} DR3 ID & RA & Dec & $l$ & $b$ & $\pi$ & $\rm G$ & $E(B-V)$ & Orbit \\
         & & & & & [deg] & [deg] & [mas] & [mag] &  \\ 
        \hline
        ra\_0103-7050\_s236 & $4691686763639138688$ & 1:03:18.24 & -70:07:30.65 & $301.45$ & $-46.97$ & $0.04 \pm 0.04$ & $16.44$ & $0.02$ & Halo \\
        ra\_0752-5047\_s47 & $5513982132481251456$ & 7:58:33.25 & -50:44:34.27 & $264.82$ & $-10.98$ & $0.11 \pm 0.04$ & $16.38$ & $0.30$ & Halo \\
        ra\_1639-2632\_s419 & $6046167303889429888$ & 16:40:04.95 & -26:37:01.99 & $353.40$ & $\phantom{-}13.13$ & $0.36 \pm 0.10$ & $17.18$ & $0.45$ & Prograde Disk \\
        ra\_1648-2642\_s91 & $6033877134724691328$ & 16:51:13.49 & -27:27:24.53 & $354.34$ & $\phantom{-}10.68$ & $0.07 \pm 0.07$ & $16.56$ & $0.21$ & Halo \\
        ra\_1659-2154\_s261 & $4126309672676244096$ & 16:58:15.46 & -21:37:03.46 & $0.07$ & $\phantom{-}12.96$ & $0.32 \pm 0.07$ & $16.97$ & $0.43$ & Prograde Disk \\
        ra\_1659-2154\_s347 & $4126178792119445888$ & 16:56:39.10 & -22:14:05.55 & $359.33$ & $\phantom{-}12.89$ & $0.16 \pm 0.08$ & $17.28$ & $0.29$ & Halo \\
        ra\_1752-4300\_s155 & $5956522780887593344$ & 17:49:24.10 & -42:38:14.33 & $348.60$ & $-7.69$ & $0.06 \pm 0.04$ & $15.63$ & $0.18$ & Prograde Disk \\        
        \hline    
    \end{tabular}
    \begin{flushleft}
        \footnotesize
        \emph{Note.} The table is available in machine-readable format in the electronic version of the paper.
    \end{flushleft}
    \label{tab:star properties}
\end{table*}

We provide a summary of spectral normalisation and processing here, but full details can be found in \citetalias{lowe_detailed_2026}. The spectra were first continuum normalised using the fully convolutional neural network \textsc{SUPPNet} \citep{rozanski_suppnet_2022}. Since multiple observations of each star were taken across many nights to reach the desired signal-to-nose (S/N), the normalised spectra were shifted using known strong absorption features (H$_\beta$ in UVB, and H$_\alpha$ in VIS) to the observation with the highest S/N. Cosmic rays were then removed by taking $3\sigma$ of the noise, assuming that any pixels above this threshold are cosmic rays. To avoid potential contamination, $\pm 1$ pixels on either side of the cosmic ray source were also rejected. The spectra were then stacked using a weighed average (by the S/N of each spectrum). Radial velocities were measured from the stacked spectra by fitting individual strong lines using Voigt functions, done separately on the two arms. For UVB, the radial velocities were measured using H$_\beta$, H$_\gamma$ and H$_\delta$. For VIS, the measurements were taken from H$_\alpha$ and the \ion{Ca}{ii} triplet. Weighted averages using the radial velocity errors as weights of each arm were taken, with the uncertainties from the weighted standard error of the mean. These can be found in Table~\ref{tab:obs log}, alongside the observing log for the P115 stars.

\begin{table}
    \centering
    \caption{Observation log for the P115 stars. The number of observations, exposure times for each observation, and S/N for both the UVB and VIS arms are provided. For stars with multiple observations, the S/N listed are the combined values for each arm (added in quadrature). Additionally, we provide the average heliocentric radial velocities for each star, derived from the stacked spectra.}
    \setlength{\tabcolsep}{3.8pt}
    \begin{tabular}{l|ccccr}
        \hline
        Star ID & $N_\mathrm{exp}$ & Exp & $S/N_{\rm UVB}$ & $S/N_{\rm VIS}$ & $RV_{\rm helio, avg}$ \\
         &  & [s] & & & [km s$^{-1}$] \\
        \hline
        ra\_0103-7050\_s236 & $2$ & $3110$ & $123.7$ & $211.8$ & $68.0 \pm 0.1$ \\
        ra\_0752-5047\_s47 & $2$ & $3110$ & $93.0$ & $164.5$ & $423.0 \pm 0.1$ \\
        ra\_1639-2632\_s419 & $4$ & $3070$ & $114.0$ & $196.9$ & $163.9 \pm 0.2$ \\
        ra\_1648-2642\_s91 & $2$ & $3110$ & $126.4$ & $193.2$ & $-113.6 \pm 0.2$ \\
        ra\_1659-2154\_s261 & $3$ & $3300$ & $112.9$ & $189.0$ & $198.9 \pm 0.2$ \\
        ra\_1659-2154\_s347 & $4$ & $3070$ & $116.8$ & $176.5$ & $-341.4 \pm 0.2$ \\
        ra\_1752-4300\_s155 & $2$ & $3090$ & $170.6$ & $277.6$ & $-43.6 \pm 0.1$ \\      
        \hline
    \end{tabular}
    \begin{flushleft}
        \footnotesize
        \emph{Note.} The table is available in machine-readable format in the electronic version of the paper.
    \end{flushleft}
    \label{tab:obs log}
\end{table}

\subsection{Data Analysis}
\label{subsec: data analysis}
Here, we summarise the derivation of stellar parameters, metallicities, chemical abundances, error analysis and upper limits, alongside orbital derivation for our P115 stars. The analysis techniques presented here are identical to the steps taken for the P113 sample, with full details found in \citetalias{lowe_detailed_2026}.

\subsubsection{Stellar Parameters}
\label{subsubsec: stellar params}
We adopted the effective temperatures ($\Teff$) and surface gravities ($\logg$) previously determined by \citet{lowe_rise_2025}. In brief, $\Teff$ values were calculated using the colour--$\Teff$ relations of \citet{casagrande_galah_2021}, while $\logg$ values were derived either directly from parallaxes for stars with a parallax significance $>3\sigma$, or through isochrone fitting otherwise.

The microturbulence velocity ($\vmic$) was determined for each star using equation (4) from \citet{buder_galah_2025}. Uncertainties in $\vmic$ were derived from error propagation using errors in $\Teff$ and $\logg$. 

The projected rotational velocity ($\vsini$) for all stars was set to $0$\,km\,s$^{-1}$. This is because at the temperatures and gravities of the sample stars, any rotation of the outer layers is likely to be undetectable at the resolution of X-Shooter. 

\subsubsection{Metallicities and chemical abundances}
\label{subsubsec: metallicitys and abundances}
The local thermodynamic equilibrium (LTE) and one-dimensional geometry spectral synthesis code \textsc{Korg} \citep{wheeler_korg_2022, wheeler_korg_2023}, was used to determine the stellar metallicities and chemical abundances. The Vienna Atomic Line Database \citep{piskunov_vald_1995, kupka_vald-2_1999, ryabchikova_major_2015} was used to generate our linelist, with the MARCS model atmospheres \citep{gustafsson_grid_2008} used within \textsc{Korg}. For our analysis, we used the \texttt{FIT\_SPECTRUM} method, which uses $\chi^2$ minimisation to fit a synthetic spectra onto the given observed data in a line-by-line analysis. Fitting windows of at least $\pm 3$\,\AA{} were used for each measurement. Since our stars are metal-poor, we set $\aFe = 0.4$ for the first iteration, then for each subsequent iteration (see below for more information), we used the weighted average of our measured $\alpha$-elements (Mg, Si, Ca and Ti). Elements that could not be measured from the spectra like O and other $\alpha$ elements, were set by $\aFe$.

\begin{figure}
    \centering
    \includegraphics[width=1\linewidth]{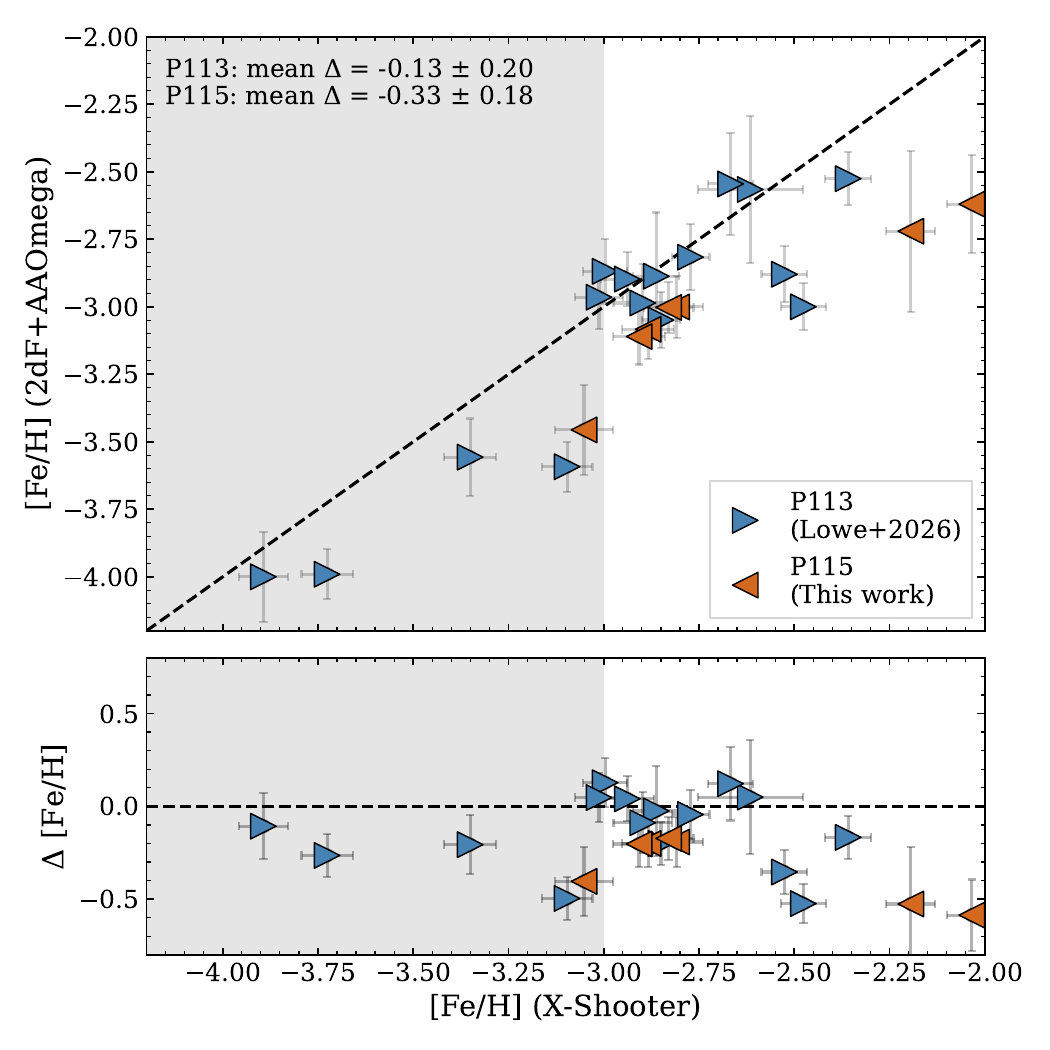}
    \caption{\textit{Upper panels}: Metallicity comparison of the values from the P115 (left-facing orange triangles; this work) and the P113 \citepalias[right-facing blue triangles;][]{lowe_detailed_2026} X-Shooter spectra, against their 2dF+AAOmega equivalents from \citet{lowe_rise_2025}. The region shaded is for all X-Shooter metallicities with $\FeH < -3.0$. \textit{Lower panels:} The difference between metallicities derived from the X-Shooter and 2dF+AAOmega spectra (2dF+AAOmega minus X-Shooter); the dashed line is for zero difference.} 
    \label{fig:xshooter metallicity}
\end{figure}

Metallicities\footnote{Here, we take metallicity to be $\FeH$.} for our sample was found by fitting all \ion{Fe}{I} lines used by \citet[see their table 3]{caffau_x-shooter_2013}, iteratively refining the initial $\FeH$ estimates from \citet{lowe_rise_2025} (detailed below). The final metallicity value was taken as the weighted average of all measured \ion{Fe}{I} lines. In Fig.~\ref{fig:xshooter metallicity}, we show that these metallicities correlate well with the 2dF+AAOmega values from \citet{lowe_rise_2025}. The mean metallicity difference (2dF+AAOmega - X-Shooter) for the P115 stars is $\Delta \FeH_{\rm P115} = -0.33 \pm 0.18$ (mean $\pm$ standard deviation), mostly driven by the two metal-rich stars at $\FeH \geq -2.25$ which have relatively large error bars on their 2dF+AAOmega values. Overall, the mean metallicity difference for these stars is consistent with that found for the P113 stars: $\Delta \FeH_{\rm P113} = -0.13 \pm 0.20$. We present the metallicities, alongside the derived $\vmic$ (see Section~\ref{subsubsec: stellar params}) and stellar parameters ($\Teff$ and $\logg$) from \citet{lowe_rise_2025} in Table~\ref{tab:star params}.

\begin{table}
    \centering
    \caption{Stellar parameters and uncertainties for our P115 stars. $\FeH$ and $\vmic$ were derived directly, whilst $\Teff$ and $\logg$ (alongside their uncertainties) were taken from \citet{lowe_rise_2025}.}
    \setlength{\tabcolsep}{2.8pt}
    \begin{tabular}{l|cccc}
        \hline
        Star ID & $\Teff$ & $\logg$ & $\FeH$ & $\vmic$ \\
         & [K] & &  & [km s$^{-1}$] \\
        \hline
        ra\_0103-7050\_s236 & $4620 \pm 10$ & $1.00 \pm 0.30$ & $-2.88 \pm 0.07$ & $2.42 \pm 0.30$ \\
        ra\_0752-5047\_s47 & $4780 \pm 90$ & $2.39 \pm 0.62$ & $-3.05 \pm 0.08$ & $1.48 \pm 0.37$ \\
        ra\_1639-2632\_s419 & $\phantom{0}5400 \pm 100$ & $2.87 \pm 0.50$ & $-2.03 \pm 0.06$ & $1.60 \pm 0.30$ \\
        ra\_1648-2642\_s91 & $5060 \pm 70$ & $2.02 \pm 0.32$ & $-2.81 \pm 0.07$ & $1.96 \pm 0.26$ \\
        ra\_1659-2154\_s261 & $\phantom{0}5400 \pm 100$ & $2.92 \pm 0.46$ & $-2.19 \pm 0.06$ & $1.58 \pm 0.28$ \\
        ra\_1659-2154\_s347 & $\phantom{0}5500 \pm 100$ & $3.08 \pm 0.36$ & $-2.83 \pm 0.07$ & $1.56 \pm 0.21$ \\
        ra\_1752-4300\_s155 & $4800 \pm 50$ & $1.41 \pm 0.21$ & $-2.91 \pm 0.07$ & $2.21 \pm 0.20$ \\
        \hline
    \end{tabular}
    \begin{flushleft}
        \footnotesize
        \emph{Note.} The table is available in machine-readable format in the electronic version of the paper.
    \end{flushleft}
    \label{tab:star params}
\end{table}

Once the metallicities were derived, abundances were then measured for the following elements: \ion{Na}{I}, \ion{Mg}{I}, \ion{Al}{I}, \ion{Si}{I}, \ion{Ca}{II}, \ion{Sc}{II}, \ion{Ti}{II}, \ion{Cr}{I}, \ion{Mn}{I}, \ion{Co}{I}, \ion{Ni}{I}, \ion{Sr}{II}, \ion{Ba}{II} and \ion{Eu}{II}. For each element, the initial guess was set at the solar scale ($\XFe{X} = 0$), then for each subsequent fit, the previously-fitted value was adopted as the initial guess. For elements with multiple lines, a weighted average across the measured abundances for each line was taken. 

C and N were also measured using the molecular bands CH ($\sim 4300$\,\AA{}) and NH ($\sim 3360$\,\AA{}), respectively. The fits to these regions can be found in Fig.~\ref{fig:cfe fits} and Fig.~\ref{fig:nfe fits}, respectively. Evolutionary mixing corrections were applied to the fitted C abundances using \citet{placco_carbon-enhanced_2014}\footnote{\url{https://vplacco.pythonanywhere.com/}}, with typical corrections ranging from $+0.00$\,dex for our dwarfs, to $+0.76$\,dex for our coolest, lowest $\logg$ giants.

Once both the metallicities and abundances were measured, the process was repeated nine times, where the difference in derived abundances at the end of the iterations was typically less than $+0.01$\,dex. For subsequent iterations after the first, the final abundance value from the previous round for the element measured was adopted as the initial guess. Additionally, all previously-determined abundances from the previous round were also set as fixed parameters in each measurement. 

All abundances reported are determined in LTE, with the exception of Ca. This is because we had to use \ion{Ca}{II} triplet lines, known to be affected by non-local thermodynamic equilibrium (NLTE), whereas the Literature Comparison sample uses \ion{Ca}{I} which is less prone to these effects \citep{osorio_accurate_2022}. Because of this, NLTE corrections for \ion{Ca}{II} from \citet{osorio_accurate_2022} were applied. 

Abundance errors and upper limits are determined as described in \citetalias{lowe_detailed_2026}. We show the median errors for our P115 sample abundances from stellar parameter perturbation ($\Teff$ by $+100$\,K, $\logg$ by $+0.3$\,dex, and $\vmic$ by $+0.3$\,km\,s$^{-1}$) in Table~\ref{tab:abund errors}, calculated by taking $\XFe{X}_{\rm perturb} - \XFe{X}$. Each perturbation is done separately for each stellar parameter: e.g., when $\Teff$ is perturbed, $\logg$ and $\vmic$ are kept fixed. Our derived errors are consistent with those from P113 (see table 3 in \citetalias{lowe_detailed_2026}).

\begin{table}
    \centering
    \caption{Median $\XH{X}$ errors for our chemical abundances and metallicities ordered by atomic number for our P115 stars. Errors were found by perturbing $\Teff$ by $+100$\,K, $\logg$ by $+0.3$\,dex, and $\vmic$ by $+0.3$\,km\,s$^{-1}$, then calculating $\XH{X}_{\rm perturb} - \XH{X}$. Errors are split between those from perturbed $\Teff$, $\logg$ and $\vmic$.}
    \begin{tabular}{c|cccc}
        \hline
        Element & Median error & Median error & Median error \\
         & ($\Teff + 100$\,K) & ($\logg + 0.3$) & ($\vmic + 0.3$\,km s$^{-1}$) \\
        \hline
        C & $+0.21$ & $-0.10$ & $-0.00$ \\
        N & $+0.18$ & $-0.26$ & $-0.05$ \\
        Na & $+0.13$ & $-0.09$ & $-0.10$ \\
        Mg & $+0.14$ & $-0.14$ & $-0.12$ \\
        Al & $+0.04$ & $-0.06$ & $-0.11$ \\
        Si & $+0.16$ & $-0.13$ & $-0.07$ \\
        Ca & $+0.08$ & $-0.02$ & $-0.01$ \\
        Sc & $-0.02$ & $+0.13$ & $-0.13$ \\
        Ti & $+0.02$ & $+0.05$ & $-0.13$ \\
        Cr & $+0.13$ & $-0.01$ & $-0.06$ \\
        Mn & $+0.20$ & $-0.07$ & $-0.11$ \\
        Fe & $+0.14$ & $-0.10$ & $-0.12$ \\
        Co & $+0.14$ & $-0.01$ & $-0.04$ \\
        Ni & $+0.09$ & $+0.01$ & $-0.04$ \\
        Sr & $+0.08$ & $+0.06$ & $-0.25$ \\
        Ba & $+0.06$ & $+0.07$ & $-0.09$ \\
        Eu & $-0.03$ & $+0.14$ & $+0.03$ \\   
        \hline
    \end{tabular}
    \begin{flushleft}
        \footnotesize
        \emph{Note.} The table is available in machine-readable format in the electronic version of the paper.
    \end{flushleft}
    \label{tab:abund errors}
\end{table}

\subsubsection{Orbital analysis}
\label{subsubsec: orbital analysis}
Orbital analysis for the P115 sample was done in \citet{lowe_rise_2025} using their derived 2dF+AAOmega radial velocities in the \textsc{galpy}\footnote{\url{https://github.com/jobovy/galpy}} Python code with MWPotential2014 \citep{bovy_galpy_2015}. This is identical to the orbital analysis for the P113 sample. Orbits were also computed in the same way for the literature samples discussed in Section~\ref{sec:literature comparison} using their \textit{Gaia} DR3 distances and proper motions. Radial velocities were taken directly from each respective study (if not available, then the \textit{Gaia} DR3 RV values were adopted instead).

\section{Literature Comparison}
\label{sec:literature comparison}
Alongside our two X-Shooter samples, we also used data collected from the literature. The sources include \citet{yong_most_2013}, \citet{jacobson_high-resolution_2015}, \citet{marino_keck_2019} and \citet{yong_high-resolution_2021} (collectively referred to as the `Literature Comparison' sample). All were chosen based on similar choice of high-resolution spectrographs, analysis techniques, and abundances and metallicities measured consistently with the LTE abundance analysis program MOOG \citep{sneden_carbon_1973, sobeck_abundances_2011}. RV values given in the literature samples were adopted, which were all determined by cross-correlation techniques. We state the number of stars used for each dataset (see below), but we refer the reader to the individual works for the full description of analysis techniques. We only take stars with \textit{Gaia} DR3 entries to get the stellar coordinates, proper motions, and parallaxes necessary for us to perform reliable orbit calculations. The description of our orbital analysis can be found in Section~\ref{subsubsec: orbital analysis} and in \citet{lowe_rise_2025}, and that approach was used here for the literature stars. In cases where stars had multiple entries, the most recent values of radial velocities and abundance determinations were adopted. In total, we have 315 literature stars that were used in the orbital analysis.

From the sample in \citet{yong_most_2013}, we took 26 stars from the original 190 metal-poor sample, with RV values provided by \citet{norris_most_2013}. For \citet{yong_high-resolution_2021}, we took 150 observed metal-poor stars with full abundance information. In \citet{jacobson_high-resolution_2015}, we took all of the 122 metal-poor stars observed, though we note that \citet{cordoni_exploring_2021} identified incorrect RV measurements during a particular run reported in \citet{jacobson_high-resolution_2015}. So for these stars, we instead adopt the \textit{Gaia} DR3 radial velocities for all stars from that run. In \citet{marino_keck_2019}, we took all of the 17 EMP candidates observed.

\section{X-Shooter Results}
\label{sec:xshooter results}
Here, we present the metallicity and chemical abundance trends for our P115 stars studied in this work. For more information on the results for the P113 stars, we refer the reader to \citetalias{lowe_detailed_2026}.

\subsection{Chemical abundances}
\label{subsec:chemical abundances}
\begin{figure*}
    \centering
    \includegraphics[width=1.0\linewidth]{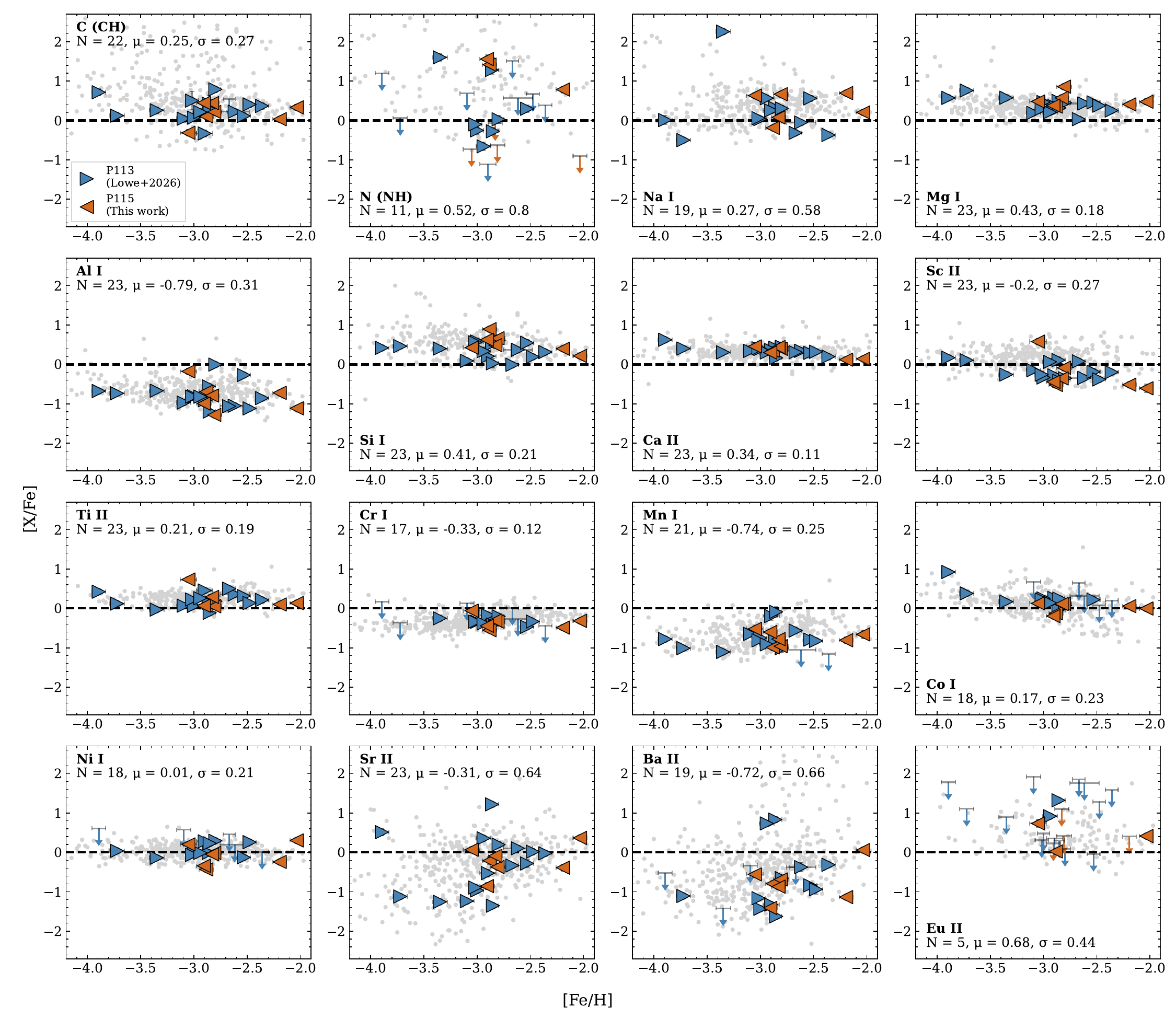}
    \caption{Chemical abundances for both P113 and P115 X-Shooter samples compared with the high-resolution Literature Comparison values in grey \citep{yong_most_2013, jacobson_high-resolution_2015, marino_keck_2019, yong_high-resolution_2021}. P113 stars are shown by right-facing blue triangle symbols, whilst P115 stars are shown by the left-facing orange triangle symbols. Each panel represents a different element measured from C (CH) to \ion{Eu}{II}. For each element, the number of X-Shooter stars with measurements, alongside their mean and standard deviations, are given. For C, the abundance measurements have been evolutionary-corrected (to be consistent with literature values). For \ion{Ca}{II}, we show the NLTE-corrected abundances, while the literature values were measured from \ion{Ca}{I} (not NLTE-corrected).}
    \label{fig:xfe abunds}
\end{figure*}

\begin{figure*}
    \centering
    \includegraphics[width=1\linewidth]{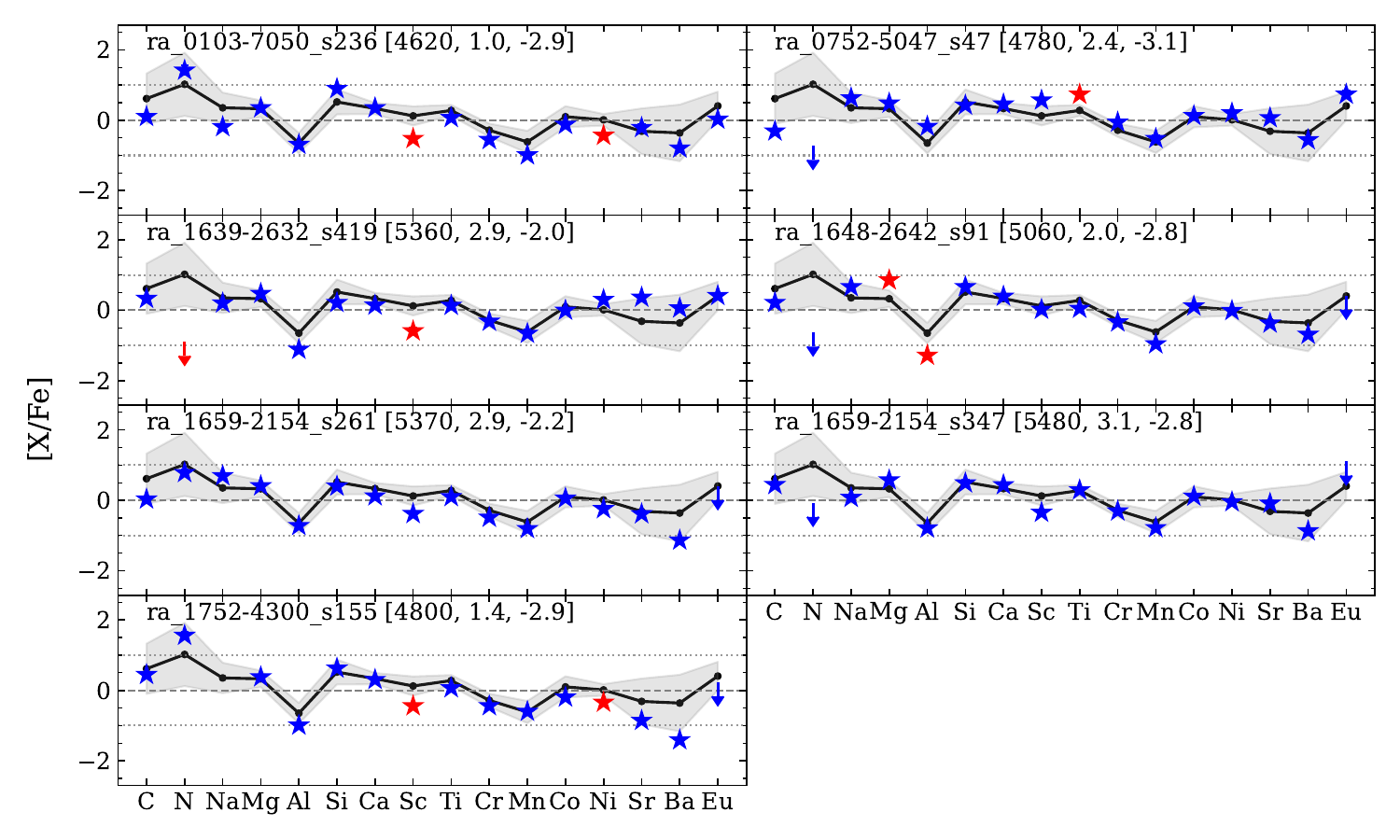}
    \caption{$\XFe{X}$ abundance trends for our seven P115 metal-poor stars across the $16$ measured elements. The panels refers to each individual star's abundance trend, with its $\Teff$, $\logg$ and $\FeH$ values indicated in the panel. For each panel, the black dots refer to the mean abundance value of each element from the Literature Comparison sample, with the grey shaded region representing its standard deviation. The star's abundance measurements are shown by the coloured star symbols. Blue symbols indicates the measurement is within $2\sigma$ of the mean Literature Comparison value, whilst red indicates otherwise. For measurements with upper-limits, a downward-facing arrow is represented instead. If greater than $2\sigma$ from the mean literature value, this is coloured red. Otherwise they are blue. The horizontal grey dashed lines are given at $\XFe{X} = 0$ and $\XFe{X} \pm 1.0$ for guidance.} 
    \label{fig:xshooter abund trend}
\end{figure*}

The chemical abundances for our P115 stars (alongside the P113 sample) are shown in Fig.~\ref{fig:xfe abunds} for our measured elements. To ensure consistency with the Literature Comparison, we report 1D LTE measurements for all elements excluding \ion{Ca}{II}, which was NLTE-corrected to be compatible with the Literature Comparison's LTE \ion{Ca}{I} abundances. The $\XFe{X}$ abundances for the P115 stars are listed in Table~\ref{tab:abunds table}. Additionally, in Fig.~\ref{fig:xshooter abund trend}, we show the abundance trends for our P115 stars against the mean and standard deviation of the Literature Comparison sample. Overall, the P115 results from Fig.~\ref{fig:xfe abunds} are consistent with both the P113 and Literature Comparison samples across all $16$ elements measured. 

\begin{table*}
    \centering
    \caption{Metallicity and abundance measurements for our P115 sample. We include the kinematic groupings beneath the star name in parenthesis. Those with no detections have their upper-limits represented instead. For C, we show both the raw ($\XFe{C}_{\rm raw}$) and evolutionary-corrected ($\XFe{C}_{\rm corr}$) values. Note that star ra\_1752-4300\_s155 has no $\XFe{Na}$ measurements due to contamination from interstellar Na absorption.}
    \setlength{\tabcolsep}{2.5pt}
    \begin{tabular}{l|ccccccccc}
        \hline
        Star & \makecell{$\FeH$ \\ $\XFe{Sc}$} 
             & \makecell{$\XFe{C}_{\rm raw}$ \\ $\XFe{Ti}$} 
             & \makecell{$\XFe{C}_{\rm corr}$ \\ $\XFe{Cr}$} 
             & \makecell{$\XFe{N}$ \\ $\XFe{Mn}$} 
             & \makecell{$\XFe{Na}$ \\ $\XFe{Co}$} 
             & \makecell{$\XFe{Mg}$ \\ $\XFe{Ni}$} 
             & \makecell{$\XFe{Al}$ \\ $\XFe{Sr}$} 
             & \makecell{$\XFe{Si}$ \\ $\XFe{Ba}$} 
             & \makecell{$\XFe{Ca}$ \\ $\XFe{Eu}$} \\
        \hline
        \makecell{ra\_0103-7050\_s236 \\ (\textit{Halo})} 
            & \makecell{$-2.88 \pm 0.07$ \\ $-0.52 \pm 0.05$} 
            & \makecell{$-0.66 \pm 0.08$ \\ $+0.07 \pm 0.03$} 
            & \makecell{$+0.10 \pm 0.08$ \\ $-0.55 \pm 0.05$} 
            & \makecell{$+1.42 \pm 0.13$ \\ $-0.99 \pm 0.07$} 
            & \makecell{$-0.19 \pm 0.06$ \\ $-0.12 \pm 0.05$} 
            & \makecell{$+0.35 \pm 0.07$ \\ $-0.43 \pm 0.05$} 
            & \makecell{$-0.69 \pm 0.06$ \\ $-0.21 \pm 0.09$} 
            & \makecell{$+0.90 \pm 0.05$ \\ $-0.79 \pm 0.04$} 
            & \makecell{$+0.35 \pm 0.02$ \\ $+0.02 \pm 0.06$} \\
        \makecell{ra\_0752-5047\_s47 \\ (\textit{Halo})}
            & \makecell{$-3.05 \pm 0.08$ \\ $+0.57 \pm 0.07$} 
            & \makecell{$-0.31 \pm 0.07$ \\ $+0.73 \pm 0.04$} 
            & \makecell{$-0.31 \pm 0.07$ \\ $-0.06 \pm 0.06$} 
            & \makecell{$< -0.73$ \\ $-0.52 \pm 0.10$} 
            & \makecell{$+0.63 \pm 0.07$ \\ $+0.13 \pm 0.05$} 
            & \makecell{$+0.48 \pm 0.07$ \\ $+0.20 \pm 0.04$} 
            & \makecell{$-0.18 \pm 0.08$ \\ $+0.07 \pm 0.07$} 
            & \makecell{$+0.43 \pm 0.08$ \\ $-0.56 \pm 0.05$}
            & \makecell{$+0.46 \pm 0.03$ \\ $+0.73 \pm 0.10$} \\
        \makecell{ra\_1639-2632\_s419 \\ (\textit{Prograde Disk})} 
            & \makecell{$-2.03 \pm 0.06$ \\ $-0.59 \pm 0.09$} 
            & \makecell{$+0.32 \pm 0.07$ \\ $+0.14 \pm 0.04$} 
            & \makecell{$+0.33 \pm 0.07$ \\ $-0.32 \pm 0.06$} 
            & \makecell{$< -0.90$ \\ $-0.66 \pm 0.10$} 
            & \makecell{$+0.20 \pm 0.06$ \\ $-0.00 \pm 0.05$} 
            & \makecell{$+0.48 \pm 0.07$ \\ $+0.30 \pm 0.06$} 
            & \makecell{$-1.11 \pm 0.04$ \\ $+0.37 \pm 0.04$} 
            & \makecell{$+0.22 \pm 0.07$ \\ $+0.06 \pm 0.06$} 
            & \makecell{$+0.14 \pm 0.03$ \\ $+0.41 \pm 0.04$} \\
        \makecell{ra\_1648-2642\_s91 \\ (\textit{Halo})} 
            & \makecell{$-2.81 \pm 0.07$ \\ $+0.02 \pm 0.07$} 
            & \makecell{$+0.19 \pm 0.08$ \\ $+0.05 \pm 0.05$} 
            & \makecell{$+0.22 \pm 0.08$ \\ $-0.34 \pm 0.05$} 
            & \makecell{$< -0.63$ \\ $-0.96 \pm 0.07$} 
            & \makecell{$+0.67 \pm 0.07$ \\ $+0.12 \pm 0.05$} 
            & \makecell{$+0.86 \pm 0.07$ \\ $-0.01 \pm 0.03$} 
            & \makecell{$-1.28 \pm 0.03$ \\ $-0.37 \pm 0.10$}
            & \makecell{$+0.67 \pm 0.08$ \\ $-0.69 \pm 0.05$} 
            & \makecell{$+0.40 \pm 0.03$ \\ $< +0.42$} \\
        \makecell{ra\_1659-2154\_s261 \\ (\textit{Prograde Disk})} 
            & \makecell{$-2.19 \pm 0.06$ \\ $-0.38 \pm 0.07$} 
            & \makecell{$+0.02 \pm 0.07$ \\ $+0.10 \pm 0.05$} 
            & \makecell{$+0.03 \pm 0.07$ \\ $-0.49 \pm 0.05$} 
            & \makecell{$+0.79 \pm 0.09$ \\ $-0.81 \pm 0.08$} 
            & \makecell{$+0.70 \pm 0.06$ \\ $+0.06 \pm 0.05$} 
            & \makecell{$+0.41 \pm 0.07$ \\ $-0.24 \pm 0.04$} 
            & \makecell{$-0.72 \pm 0.05$ \\ $-0.39 \pm 0.09$} 
            & \makecell{$+0.40 \pm 0.07$ \\ $-1.14 \pm 0.06$} 
            & \makecell{$+0.12 \pm 0.03$ \\ $< +0.40$} \\
        \makecell{ra\_1659-2154\_s347 \\ (\textit{Halo})} 
            & \makecell{$-2.83 \pm 0.07$ \\ $-0.35 \pm 0.04$} 
            & \makecell{$+0.43 \pm 0.08$ \\ $+0.29 \pm 0.05$} 
            & \makecell{$+0.44 \pm 0.08$ \\ $-0.31 \pm 0.03$} 
            & \makecell{$< -0.07$ \\ $-0.78 \pm 0.05$} 
            & \makecell{$+0.08 \pm 0.05$ \\ $+0.11 \pm 0.04$} 
            & \makecell{$+0.57 \pm 0.07$ \\ $-0.05 \pm 0.03$} 
            & \makecell{$-0.79 \pm 0.03$ \\ $-0.09 \pm 0.09$} 
            & \makecell{$+0.49 \pm 0.08$ \\ $-0.87 \pm 0.06$} 
            & \makecell{$+0.44 \pm 0.03$ \\ $< +1.10$} \\
        \makecell{ra\_1752-4300\_s155 \\ (\textit{Prograde Disk})} 
            & \makecell{$-2.91 \pm 0.07$ \\ $-0.44 \pm 0.05$} 
            & \makecell{$-0.06 \pm 0.08$ \\ $+0.07 \pm 0.04$} 
            & \makecell{$+0.45 \pm 0.08$ \\ $-0.44 \pm 0.05$} 
            & \makecell{$+1.56 \pm 0.11$ \\ $-0.60 \pm 0.07$} 
            & \makecell{ \\ $-0.19 \pm 0.05$} 
            & \makecell{$+0.38 \pm 0.07$ \\ $-0.34 \pm 0.04$} 
            & \makecell{$-0.99 \pm 0.05$ \\ $-0.86 \pm 0.09$} 
            & \makecell{$+0.63 \pm 0.05$ \\ $-1.41 \pm 0.04$} 
            & \makecell{$+0.30 \pm 0.02$ \\ $< +0.23$} \\
        \hline
    \end{tabular}
    \begin{flushleft}
        \footnotesize
        \emph{Note.} The table is available in machine-readable format in the electronic version of the paper.
    \end{flushleft}
    \label{tab:abunds table}
\end{table*}

We now report our findings for the following nucleosynthetic groups: C and N abundances (Section~\ref{subsubsec:light elements}), $\alpha$ elements (Section~\ref{subsubsec:alpha elements}) and r-process elements (Section~\ref{subsubsec:r process elements}). We note that none of the stars in our P115 sample has a measurable Li abundance.

\subsubsection{C and N abundances}
\label{subsubsec:light elements}
\begin{figure}
    \centering
    \includegraphics[width=1\linewidth]{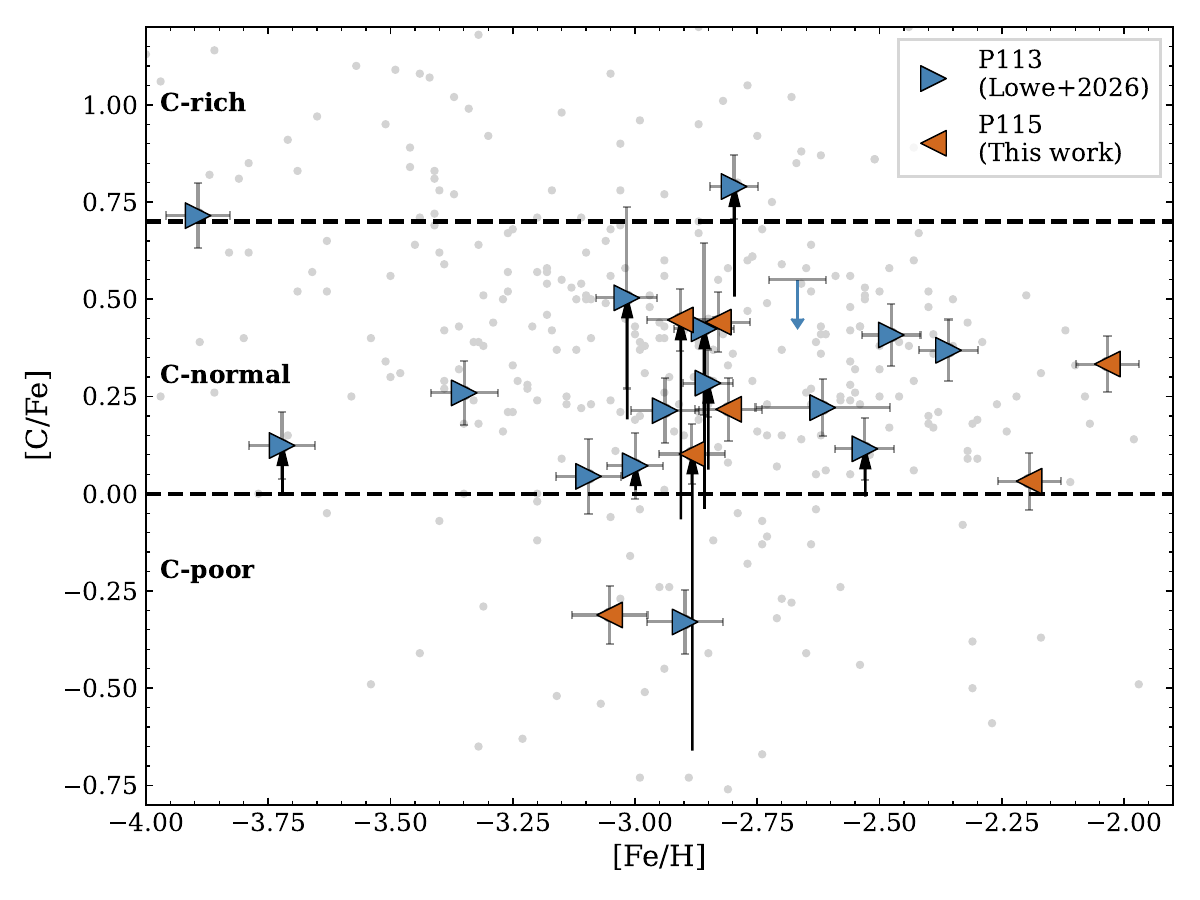}
    \caption{Evolutionary-corrected C abundances for both the P113 and P115 samples. The Literature Comparison is shown by the light grey points. Upper-limits on C are shown by downwards-facing arrows. Using the definitions from \citet{aoki_carbon-enhanced_2007}, the plot is separated into three regions: C-rich ($\XFe{C} > 0.7$), C-normal ($0 < \XFe{C} \leq 0.7$) and C-poor ($\XFe{C} \leq 0$). Stars with evolutionary corrections have their raw measurements represented by upwards-facing black arrows.} 
    \label{fig:c enhancement}
\end{figure}

Fig.~\ref{fig:c enhancement} depicts the categories of C-enhancement for the P113 and P115 samples. Here, we adopt the definitions from \citet{aoki_carbon-enhanced_2007}: C-rich ($\XFe{C} > +0.7$), C-normal ($+0.0 < \XFe{C} \leq +0.7$) and C-poor ($\XFe{C} \leq +0.0$). For our P115 sample, we have no C-rich stars, one C-poor star, with the remainder C-normal. The one interesting case is the C-poor star ra\_0752-5047\_s47, having the lowest $\XFe{C}_{\rm corrected}$ abundance for both the P115 and P113 samples at $\XFe{C}_{\rm corrected} = -0.31 \pm 0.07$. The evolutionary correction is small at $+0.01$\,dex, suggesting this star was formed out of gas that was depleted in C. As will be discussed later, this star is also enhanced in both $\alpha$ and neutron capture process elements. The implications of the abundances in this star will be explored in Section~\ref{susec:peculiar star ra0752}.

\begin{figure}
    \centering
    \includegraphics[width=1\linewidth]{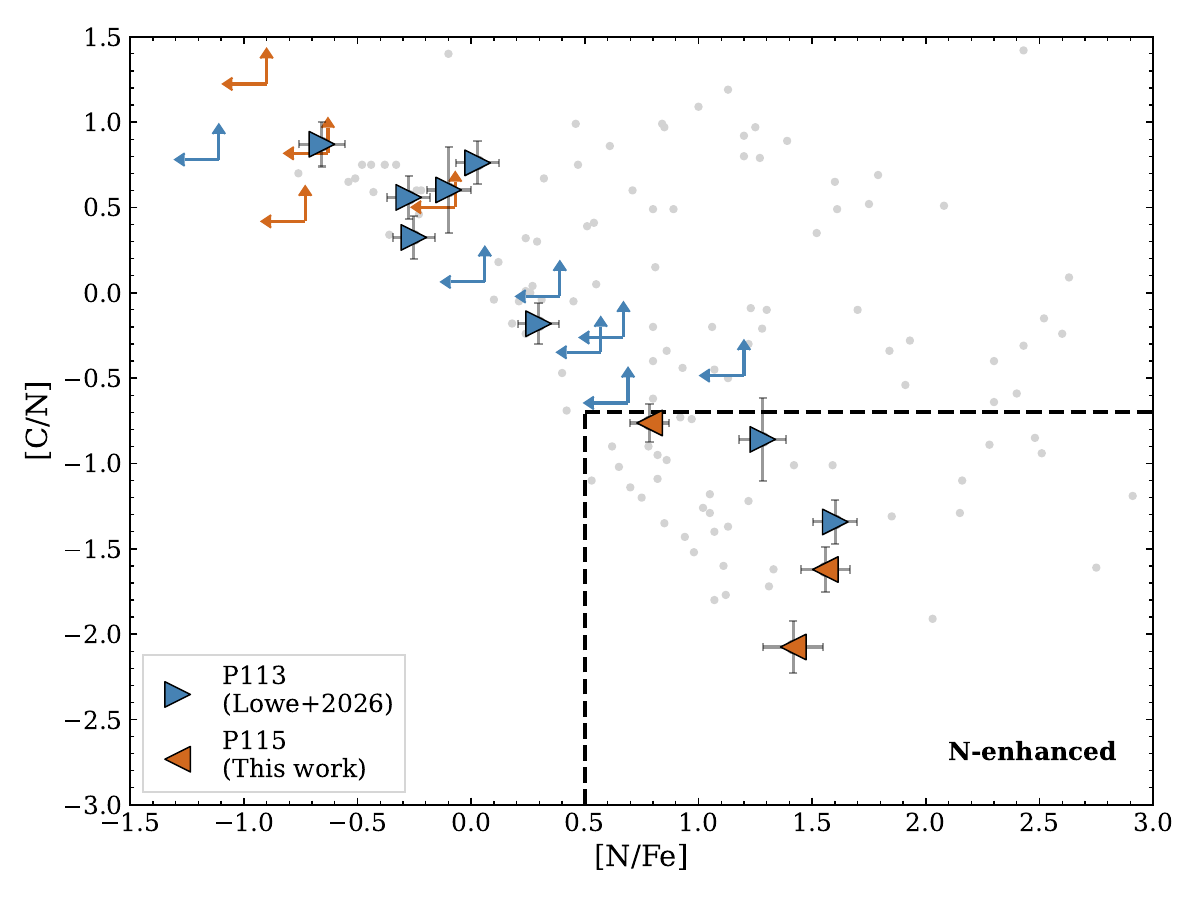}
    \caption{$\XFe{N}$ and $\XY{C}{N}$ abundances for our samples of P113 and P115 stars, with the Literature Comparison represented by the grey background points. Since no evolutionary corrections were applied to N, for consistency, we used the uncorrected C abundances when determining $\XY{C}{N}$. Stars with upper-limits in $\XFe{N}$ (represented by leftwards-facing arrows), but detections in $\XFe{C}$ have lower-limits for $\XY{C}{N}$ (represented by upwards-facing arrows). Using the NEMP definition from \citet{johnson_search_2007}, stars are assigned NEMP if $\XFe{N} > 0.5$ and $\XY{C}{N} < -0.5$, represented by the black dashed rectangle on the plot. For the P115 sample, we identify three possible NEMP stars.} 
    \label{fig:n enhancemenet}
\end{figure}

In Fig.~\ref{fig:n enhancemenet} and Table~\ref{tab:cn values}, we present $\XY{C}{N}$ as a function of $\XFe{N}$ for our data. We note that while \citet{placco_carbon-enhanced_2014} only correct for C, stellar mixing states that C depletion and N enhancement are correlated. We identified possible nitrogen enhanced metal-poor (NEMP) stars in Fig.~\ref{fig:n enhancemenet} by employing the definition provided by \citet{johnson_search_2007}: $\XFe{N} > +0.5$ and $\XY{C}{N} < -0.5$. For the P115 sample, there are three such stars (shown by the asterisk in Table~\ref{tab:cn values}). Interestingly, all three NEMP stars have $\XH{N} \approx -1.4$. Using the raw $\XFe{C}$ values, we find that the sum of C+N is enhanced over the solar value for each NEMP, with values of $[\rm{C + N/Fe}] = +0.7$, $+0.3$ and $+0.9$\,dex respectively.

We note that star ra\_1639-2632\_s419 has the lowest N upper-limit in the P115 sample at $\XFe{N} < -0.90$. This is similar to star ra\_1658-2454\_s22 from the P113 sample at $\XFe{N} < -1.11$, though unlike ra\_1658-2454\_s22, our P115 star ra\_1639-2632\_s419 is not anomalous in any other elements.

\begin{table}
    \centering
    \caption{$\XFe{C}_{\rm raw}$, $\XFe{N}$ and $\XY{C}{N}$ values for our P115 sample. Note that the uncorrected C values were used when calculating $\XY{C}{N}$. A lower-limit on $\XY{C}{N}$ was given if there was an upper-limit on $\XFe{N}$, but a detection for $\XFe{C}$. Stars with an asterisk are identified as being NEMP.}
    \begin{tabular}{l|ccc}
        \toprule
        Star & $\XFe{C}_{\rm{raw}}$ & $\XFe{N}$ & $\XY{C}{N}$ \\
        \hline
        ra\_0103-7050\_s236 * & $-0.66 \pm 0.08$ & $+1.42 \pm 0.13$ & $-2.08 \pm 0.15$ \\
        ra\_0752-5047\_s47 & $-0.31 \pm 0.07$ & $<-0.73$ & $>+0.42$ \\
        ra\_1639-2632\_s419 & $+0.32 \pm 0.07$ & $<-0.90$ & $>+1.22$ \\
        ra\_1648-2642\_s91 & $+0.19 \pm 0.08$ & $<-0.63$ & $>+0.82$ \\
        ra\_1659-2154\_s261 * & $+0.02 \pm 0.07$ & $+0.79 \pm 0.09$ & $-0.76 \pm 0.11$ \\
        ra\_1659-2154\_s347 & $+0.43 \pm 0.08$ & $<-0.07$ & $>+0.50$ \\
        ra\_1752-4300\_s155 * & $-0.06 \pm 0.08$ & $+1.56 \pm 0.11$ & $-1.62 \pm 0.13$ \\       
        \hline
    \end{tabular}
    \begin{flushleft}
        \footnotesize
        \emph{Note.} The table is available in machine-readable format in the electronic version of the paper.
    \end{flushleft}
    \label{tab:cn values}
\end{table}

\subsubsection{$\alpha$ elements}
\label{subsubsec:alpha elements}
\begin{figure}
    \centering
    \includegraphics[width=1\linewidth]{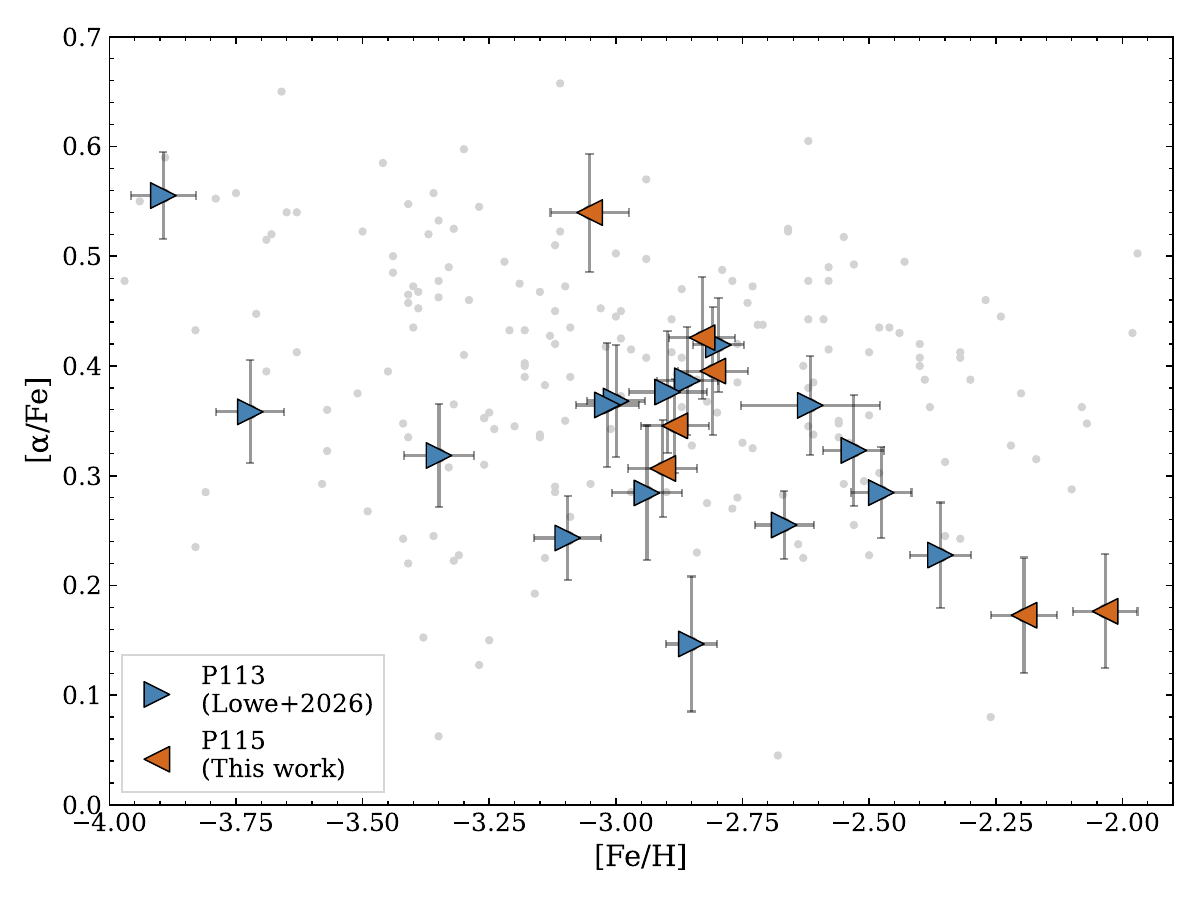}
    \caption{$\aFe$ abundances for the P113 and P115 stars, alongside the Literature Comparison (in grey). We took $\aFe$ as the weighted mean of our measured $\alpha$-elements: Ca, Mg, Sc and Ti. The errors are taken as the average of the individual elemental errors. For consistency, the Literature Comparison $\aFe$ abundances were derived using the same elements.} 
    \label{fig:alpha enhancemenet}
\end{figure}

The $\alpha$-elements, comprising here of Mg, Si, Ca and Ti, are primarily created through stellar nucleosynthesis, then ejected through core-collapse supernovae (Type II). We took $\aFe$ to be the weighted mean of our four measured $\alpha$-elements. To account for correlated systematic uncertainties in our abundances, the final errors on $\aFe$ were taken as the average of the individual elemental errors. These abundances are shown in Fig.~\ref{fig:alpha enhancemenet} and Table~\ref{tab:alpha values}. 

\begin{table}
    \centering
    \caption{$\rm {[\alpha/Fe]}$ values for the P115 sample, derived from the weighted mean of Ca, Mg, Si and Ti abundances.}
    \begin{tabular}{l|cc}
        \toprule
        Star & $\FeH$  & $\rm {[\alpha/Fe]}$ \\
        \hline
        ra\_0103-7050\_s236 & $-2.88 \pm 0.07$ & $+0.35 \pm 0.04$ \\
        ra\_0752-5047\_s47 & $-3.05 \pm 0.08$ & $+0.54 \pm 0.05$ \\
        ra\_1639-2632\_s419 & $-2.03 \pm 0.06$ & $+0.18 \pm 0.05$ \\
        ra\_1648-2642\_s91 & $-2.81 \pm 0.07$ & $+0.40 \pm 0.06$ \\
        ra\_1659-2154\_s261 & $-2.19 \pm 0.06$ & $+0.17 \pm 0.05$ \\
        ra\_1659-2154\_s347 & $-2.83 \pm 0.07$ & $+0.43 \pm 0.06$ \\
        ra\_1752-4300\_s155 & $-2.91 \pm 0.07$ & $+0.31 \pm 0.04$ \\
        \hline
    \end{tabular}
    \begin{flushleft}
        \footnotesize
        \emph{Note.} The table is available in machine-readable format in the electronic version of the paper.
    \end{flushleft}
    \label{tab:alpha values}
\end{table}

From Fig.~\ref{fig:alpha enhancemenet} we see that the P115 star ra\_0752-5047\_s47 has, at $\aFe = +0.54 \pm 0.02$, the largest $\alpha$-enhancement in this sample. The value is mostly driven by the high Ti abundance at $\XFe{Ti} = +0.73 \pm 0.04$, with the other $\alpha$-elemental abundances not as enhanced, but still relatively high ($\XFe{Mg} = +0.48 \pm 0.07$, $\XFe{Si} = +0.43 \pm 0.08$, $\XFe{Ca} = +0.46 \pm 0.06$). Interestingly, this star is also C-poor (see Section~\ref{subsubsec:light elements}), and as will be discussed in Section~\ref{subsubsec:r process elements}, is also enhanced in neutron capture elements. Further details of this star can be found in Section~\ref{susec:peculiar star ra0752}.

\subsubsection{Neutron capture elements}
\label{subsubsec:r process elements}
The neutron capture elements measured in this work are Sr, Ba and Eu, which are produced by the slow neutron capture process (s-process), or by the rapid neutron capture process (r-process). Of the stars in Table~\ref{tab:baeu values}, which lists $\XFe{Eu}$, $\XY{Ba}{Eu}$ and r-I status for the P115 sample, two stars meet the definition of r-I (moderately r-process enhanced; $+0.3 \leq \XFe{Eu} \leq +1.0$ and $\XY{Ba}{Eu} < +0.0$), and none meet the definition of r-II (strongly r-process enhanced; $\XFe{Eu} > +1.0$ and $\XY{Ba}{Eu} < +0.0$) \citep{christlieb_hamburgeso_2004}.

\begin{table}
    \centering
    \caption{$\XFe{Eu}$ and $\XY{Ba}{Eu}$ values for our P115 sample. Those with detections in Ba, but upper-limits in Eu were given a lower-limit in $\XY{Ba}{Eu}$. Upper-limits in both Ba and Eu were not included. Those identified as r-I are indicated at the end of the table with `Y'. No stars in the sample identify with r-II.}
    \begin{tabular}{l|ccc}
        \toprule
        Star & $\XFe{Eu}$ & $\XY{Ba}{Eu}$ & r-I status \\
         & & & (Y/N) \\
        \hline
        ra\_0103-7050\_s236 & $+0.02 \pm 0.06$ & $-0.82 \pm 0.08$ & N \\
        ra\_0752-5047\_s47 & $+0.73 \pm 0.10$ & $-1.29 \pm 0.11$ & Y \\
        ra\_1639-2632\_s419 & $+0.41 \pm 0.04$ & $-0.35 \pm 0.07$ & Y\\
        ra\_1648-2642\_s91 & $<+0.42$ & $>-1.10$ & N \\
        ra\_1659-2154\_s261 & $<+0.40$ & $>-1.54$ & N \\
        ra\_1659-2154\_s347 & $<+1.10$ & $>-1.98$ & N \\
        ra\_1752-4300\_s155 & $<+0.23$ & $>-1.64$ & N \\        
        \hline
    \end{tabular}
    \begin{flushleft}
        \footnotesize
        \emph{Note.} The table is available in machine-readable format in the electronic version of the paper.
    \end{flushleft}
    \label{tab:baeu values}
\end{table}

\begin{figure}
    \centering
    \includegraphics[width=1\linewidth]{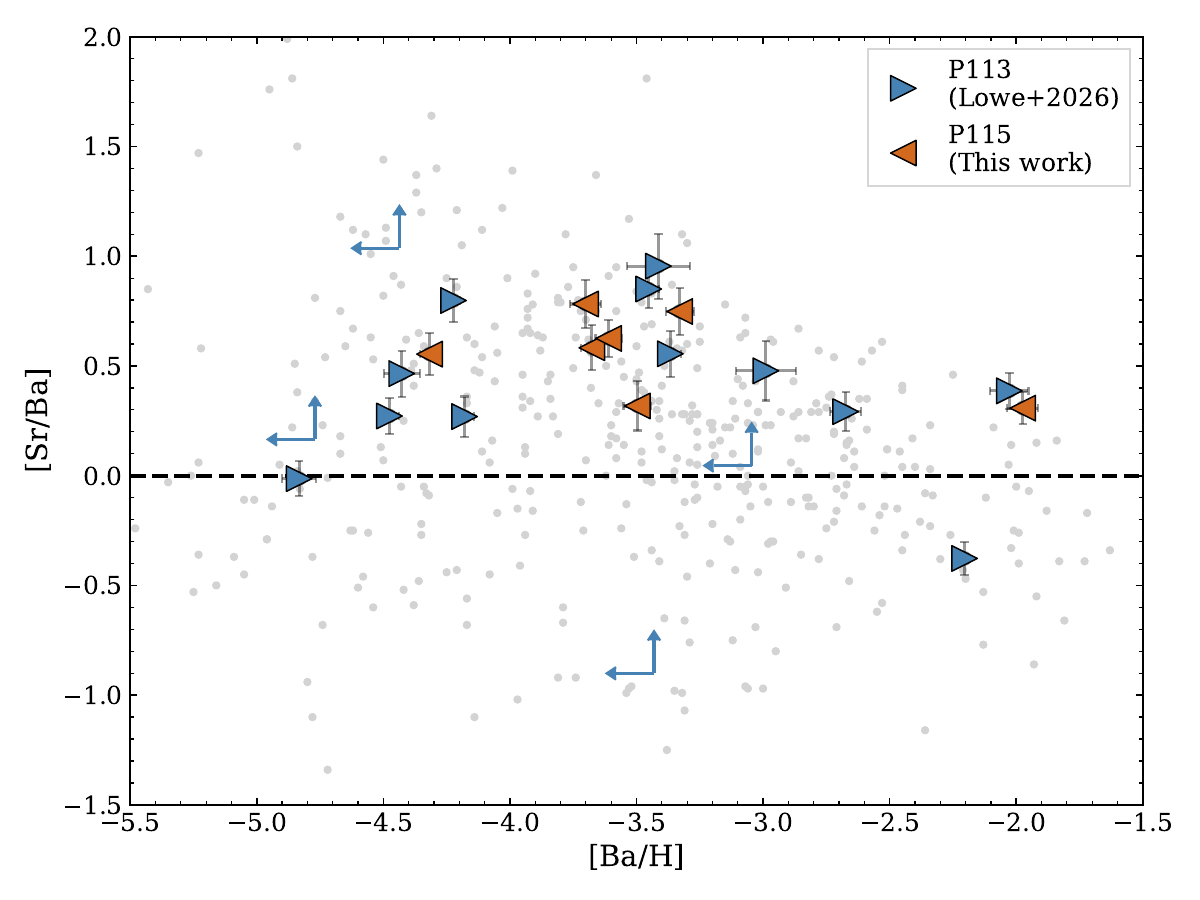}
    \caption{$\XY{Sr}{Ba}$ versus $\XY{Ba}{H}$ plot for both the P113 and P115 samples, alongside with Comparison Literature in the background (grey). Stars with upper-limits in $\XY{Ba}{H}$ (represented by leftward-facing arrows) have a corresponding lower-limit in $\XY{Sr}{Ba}$ (represented by upward-facing arrows).}
    \label{fig:srba plot}
\end{figure}

Fig.~\ref{fig:srba plot} depicts $\XY{Sr}{Ba}$ versus $\XH{Ba}$ for the P115 stars. According to \citet{sitnova_unlocking_2025} (using the models from \citet{rizzuti_explaining_2025}), the r-process yields $\XY{Ba}{Eu} = -0.87$ and $\XY{Sr}{Ba} = -0.31$. Both our r-I stars are inconsistent with these values, suggesting mixture of r- and s- process contributions.

\subsection{The peculiar star ra\_0752-5047\_s47}
\label{susec:peculiar star ra0752}
The chemical abundance pattern of star ra\_0752-5047\_s47 ($\FeH = -3.05 \pm 0.08$) is characterised by its depleted C abundances ($\XFe{C} = -0.31$ with a small evolutionary correction of $+0.01$\,dex), strong Ti abundance at $\XFe{Ti} = +0.73 \pm 0.04$ with an $\alpha$ abundance of $\aFe = +0.54 \pm 0.05$, moderately enhanced Sc abundance at $\XFe{Sc} = +0.57 \pm 0.07$, and moderate neutron capture abundance enhancement classifying it as r-I star ($\XFe{Ba} = -0.56 \pm 0.05$, $\XFe{Eu} = +0.73 \pm 0.10$ and $\XY{Ba}{Eu} = -1.29 \pm 0.11$). The star is similar to the P113 star, ra\_1633-2814\_s284, which has an $\alpha$ abundance of $\aFe = +0.56 \pm 0.02$, although with slightly enhanced Ca instead of Ti, and is not enhanced in neutron-capture elements.

In general, metal-poor stars with $\alpha$-enhancement result from contributions primarily from Type II supernova, that produce higher amounts of $\alpha$-elements relative to Fe \citep[e.g.][]{spite_nucleosynthesis_1978, tinsley_evolution_1980, kobayashi_origin_2020}. Given this star has strong Ti and moderate Sc enhancements, it is possible that the abundances resulted from a progenitor that ended its life as a jet-induced hypernova. Such events can explain the enhanced $\XFe{Ti}$ and $\XFe{Sc}$, and, coupled with ``no fallback'' could also lead to the depleted $\XFe{C}$ value. A separate r-process event is then needed explain the r-I enhancement. However, jet-induced hypernovae also under-produce Na and Mg, which is not the case for our star, where instead we see normal Na and Mg abundances. Therefore, future work with high-resolution spectroscopy is required to investigate this possibility further. A determination of the Zn abundance is particularly important, as an over-abundance is characteristic of large explosion energy supernovae (hypernovae) \citep{nomoto_nucleosynthesis_2013}.

\section{Chemodynamic Results}
\label{sec:chemodynamic results}
Here, we show the chemodynamic results of our combined sample of the P113, P115 and Literature Comparison stars (with \textit{Gaia} DR3 entries). In Section~\ref{subsec:orbit selection} we derive orbits to separate the combined sample into different kinematic classifications, then in Section~\ref{subsec:chemodynamics results}, we identify (if any) abundance trends based on these groupings.

\subsection{Orbital selection}
\label{subsec:orbit selection}
\begin{figure}
    \centering
    \includegraphics[width=1\linewidth]{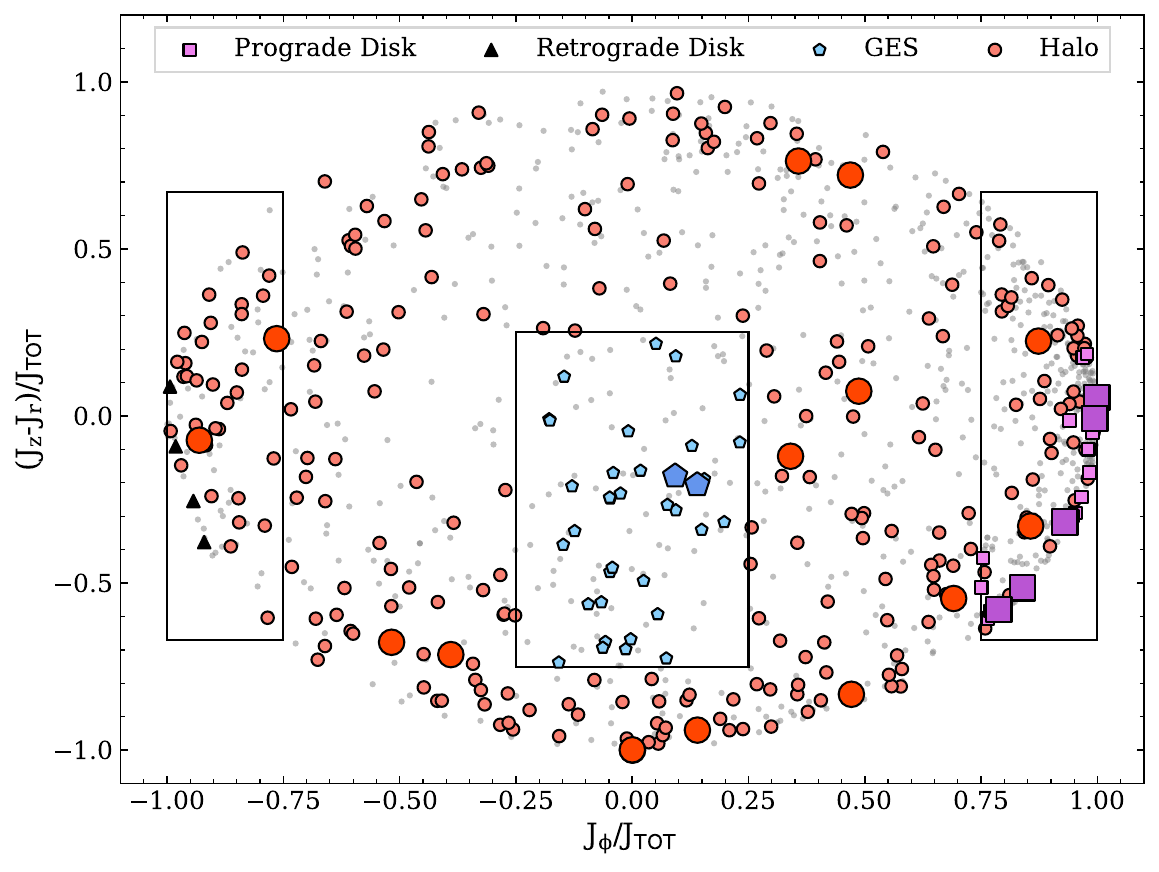}
    \caption{Kinematic selection of the combined X-Shooter (large symbols) and Literature Comparison (small symbols) samples in action momentum space. Stars from \citet{lowe_rise_2025} with $\FeH \leq -2.0$ are shown in the background as light grey points. The x-axis, $J_\phi/J_{\rm TOT}$, represents the azimuthal component of the action vector. The y-axis, $(J_z - J_r)/J_{\rm TOT}$, represents the difference between the star's vertical and radial actions. Both axes are normalised by the total angular momentum, $J_{\rm TOT}$. See text for further details on the adopted classification.} 
    \label{fig:circle plot}
\end{figure}

Using the techniques outlined in Section~\ref{subsubsec: orbital analysis}, we calculated orbits for the combined X-Shooter \citep[done in][]{lowe_rise_2025} and Literature Comparison samples with the MWPotential2014 Galactic potential. With our derived orbital parameters, we adopted the classifications defined in \citet{sestito_exploring_2021} to separate our data into the prograde disk, retrograde disk, GSE and halo kinematic groupings based on their combination of azimuthal ($J_\phi$), vertical ($J_z$) and radial ($J_r$) action vector components (normalised by the total angular momentum, $J_{\rm TOT} = \sqrt{J_\phi^2 + J_z^2 + J_r^2}$). For the prograde disk, stars are assigned to this group if $J_\phi/J_{\rm TOT} \geq 0.75$ and $z_{\rm max} < 3$\,kpc. Similarly, stars are assigned to the retrograde disk if $J_\phi/J_{\rm TOT} \leq -0.75$ and $z_{\rm max} < 3$\,kpc. If stars meet the $J_\phi/J_{\rm TOT}$ criteria, but not the $z_{\rm max} < 3$\,kpc requirement for either disk population, then they are assigned as halo stars. For the GSE, we adopt the classification defined by \citet{belokurov_co-formation_2018} and \citet{helmi_merger_2018}, whereby $-0.25 \leq J_\phi/J_{\rm TOT} \leq 0.75$ and $-0.75 \leq (J_z - J_r)/J_{\rm TOT} \leq 0.25$. All remaining stars are classified as halo. The classifications are shown in Fig.~\ref{fig:circle plot} and the number of stars in each grouping is given in Table~\ref{tab:orbit classification}.

For the X-Shooter sample, $26.1$\% of the stars exhibit prograde disk orbits, compared to only $5.1$\% in the Literature Comparison. This overdensity is not due to a kinematic pre-selection, but rather an observational footprint effect: the \citet{lowe_rise_2025} targets were preferentially observed in fields near the Galactic plane. In contrast, the high-resolution literature stars were followed-up from low-resolution studies \citep[e.g.,][]{christlieb_hamburgeso_2004, christlieb_stellar_2008, schorck_stellar_2009, dacosta_skymapper_2019} that surveyed high-latitude regions to maximise the discovery of metal-poor halo candidates. Our prograde disk sample covers an eccentricity ($e$) range of $0 \leq e \leq 0.8$, with the maximum Galactocentric radii ($r_{\rm max}$) covering the range $5 \leq r_{\rm max} \leq 16$\,kpc.

\begin{table}
    \centering
    \caption{Kinematic classifications for the X-Shooter (both P113 and P115), Literature Comparison and combined dataset samples. Each entry states the number of stars in that orbit for a given dataset, followed by its percentage.}
    \setlength{\tabcolsep}{3.2pt}
    \begin{tabular}{l|ccc}
        \hline
        Orbit & X-Shooter & Literature Comparison & All \\
         & (23 stars) & (315 stars) & (338 stars) \\
        \hline
        Prograde Disk & $6 (26.1\%)$ & $16 (5.1\%)$ & $22 (6.5\%)$ \\
        Retrograde Disk & $0 (0.0\%)$ & $4 (1.3\%)$ & $4 (1.2\%)$ \\
        GSE & $2 (8.7\%)$ & $34 (10.8\%)$ & $36 (10.7\%)$ \\
        Halo & $15 (65.2\%)$ & $261 (82.9\%)$ & $276 (81.7\%)$ \\
        \hline
    \end{tabular}
    \begin{flushleft}
        \footnotesize
        \emph{Note.} The table is available in machine-readable format in the electronic version of the paper.
    \end{flushleft}
    \label{tab:orbit classification}
\end{table}

\subsection{Chemodynamics of kinematic groupings}
\label{subsec:chemodynamics results}
To investigate the chemodynamical properties of our sample, we combined the chemical abundances determined in Section~\ref{subsec:chemical abundances} with the orbital classifications outlined in Section~\ref{subsec:orbit selection}, partitioning the stars into the prograde disk, halo, and GSE populations. Due to low-number statistics, we excluded the retrograde disk stars from our analysis. 

For each elemental species from C to Eu (excluding N due to low number of measurements in both the X-Shooter and in the Literature Comparison samples with reliable uncertainties), we performed a linear fit of $\XFe{X}$ abundance against metallicity across the metallicity range $-4.2 \leq \FeH \leq -1.9$ for the prograde disk, GSE, and halo populations. To better gauge the true abundance uncertainties and ensure realistic observational error bars, the individual measurement errors and the intrinsic scatter of the combined data were added in quadrature to weight the fits. We then computed the $\Delta \chi^2$ statistic for each kinematic population and $\XFe{X}$ abundance: for each element, we calculated the goodness of fit for the prograde disk and GSE populations compared to the halo fit in $\XFe{X}$-$\FeH$, and report this as $\Delta \chi^2_{\rm halo}$. We consider the significance level of our final $\Delta \chi^2 =\chi^2-\chi^2_{\rm halo}$ statistic. This tells us for each element whether the fit for the halo is statistically identical to that determined by each kinematic population instead. We show the resulting $\Delta \chi^2$ values for all our elements in Table~\ref{tab:chemodynamic table}, with the linear fits shown in Appendix \ref{appendix: delta chi2}. 

\begin{table}
    \centering
    \caption{$\Delta \chi^2$ values and $\sigma$ confidence intervals for each element (excluding N) are provided for the prograde disk and GSE regions across metallicities $-4.2 \leq \FeH \leq -1.9$. The halo was used as the null hypothesis to test if the prograde disk or GSE are chemically distinct from the halo. High $\Delta \chi^2$ values with $\geq3\sigma$ confidence levels indicate that the region is chemically distinct from the halo. Lower $\Delta \chi^2$ values with $<3\sigma$ confidence levels instead indicates that the region is chemically similar to the halo. The retrograde disk was excluded due to the lack of stars potentially driving inaccurate results. N was excluded for the same reason. The plots showing the linear fits can be found in Appendix \ref{appendix: delta chi2}.}
    \begin{tabular}{c|cc|cc}
        \hline
        & \multicolumn{2}{c|}{Prograde Disk} & \multicolumn{2}{c}{GSE} \\
        Element & $\Delta \chi^2$ & $\sigma$ & $\Delta \chi^2$ & $\sigma$ \\
        \hline
        C & $0.07$ & $0.0$ & $6.50$ & $2.1$ \\
        Na & $0.19$ & $0.1$ & $1.23$ & $0.6$ \\
        Mg & $14.22$ & $3.3$ & $0.44$ & $0.2$ \\
        Al & $6.92$ & $2.2$ & $0.01$ & $0.0$ \\
        Si & $5.98$ & $2.0$ & $1.03$ & $0.5$ \\
        Ca & $2.85$ & $1.2$ & $2.29$ & $1.0$ \\
        Sc & $29.50$ & $5.1$ & $3.08$ & $1.2$ \\
        Ti & $6.25$ & $2.0$ & $0.05$ & $0.0$ \\
        Cr & $3.24$ & $1.3$ & $0.57$ & $0.3$ \\
        Mn & $0.58$ & $0.3$ & $6.34$ & $2.0$ \\
        Co & $5.22$ & $1.8$ & $2.11$ & $0.9$ \\
        Ni & $1.08$ & $0.5$ & $1.31$ & $0.6$ \\
        Sr & $2.72$ & $1.1$ & $6.65$ & $2.1$ \\
        Ba & $1.88$ & $0.9$ & $1.80$ & $0.8$ \\
        Eu & $1.75$ & $0.8$ & $5.55$ & $1.9$ \\     
        \hline
    \end{tabular}
    \begin{flushleft}
        \footnotesize
        \emph{Note.} The table is available in machine-readable format in the electronic version of the paper.
    \end{flushleft}
    \label{tab:chemodynamic table}
\end{table}

From Table~\ref{tab:chemodynamic table}, we see that the GSE population has low $\Delta \chi^2$ values across all 15 elements compared to the halo population, with no element indicating a difference at a statistical confidence level greater than $3\sigma$. This indicates that the GSE and halo have similar abundance trends. 

\begin{figure*}
    \centering
    \includegraphics[width=1\linewidth]{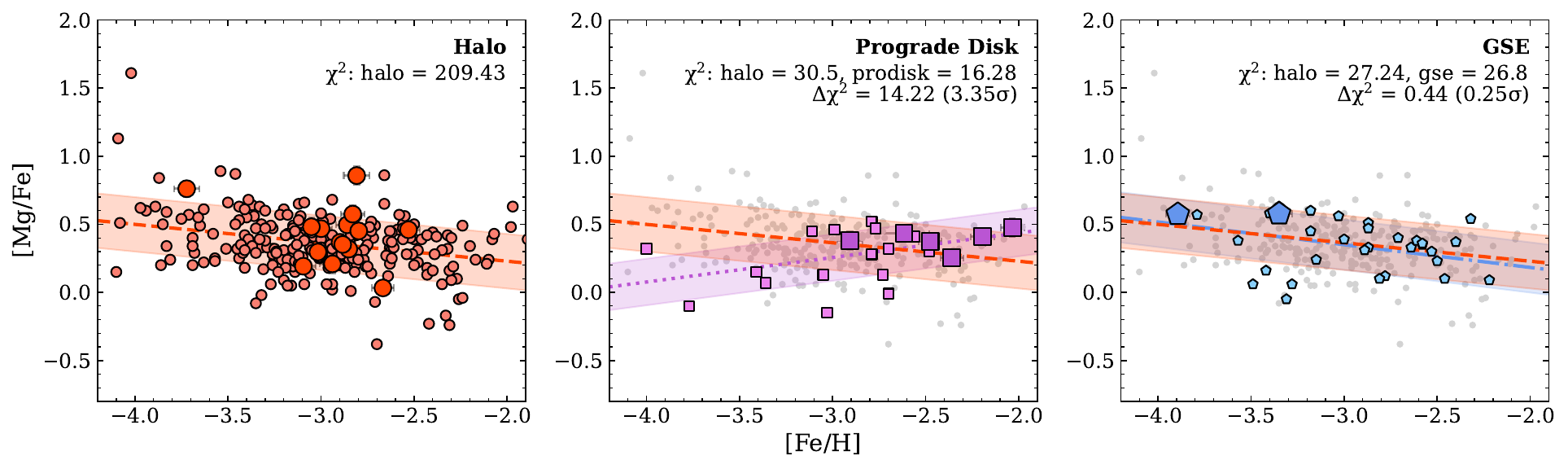}
    \includegraphics[width=1\linewidth]{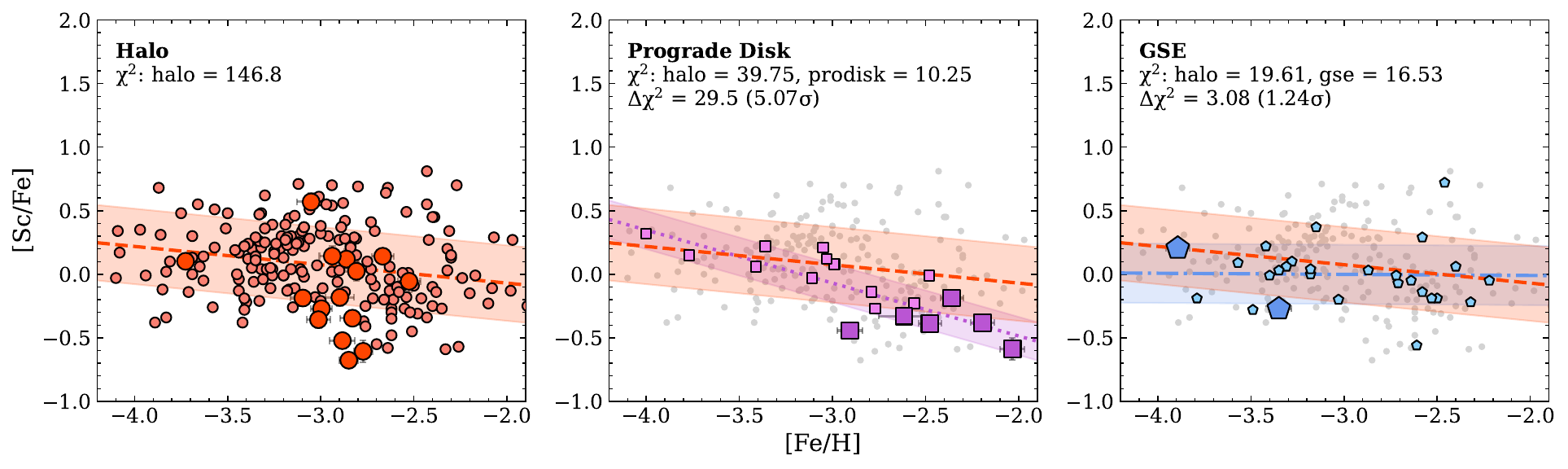}
    \includegraphics[width=1\linewidth]{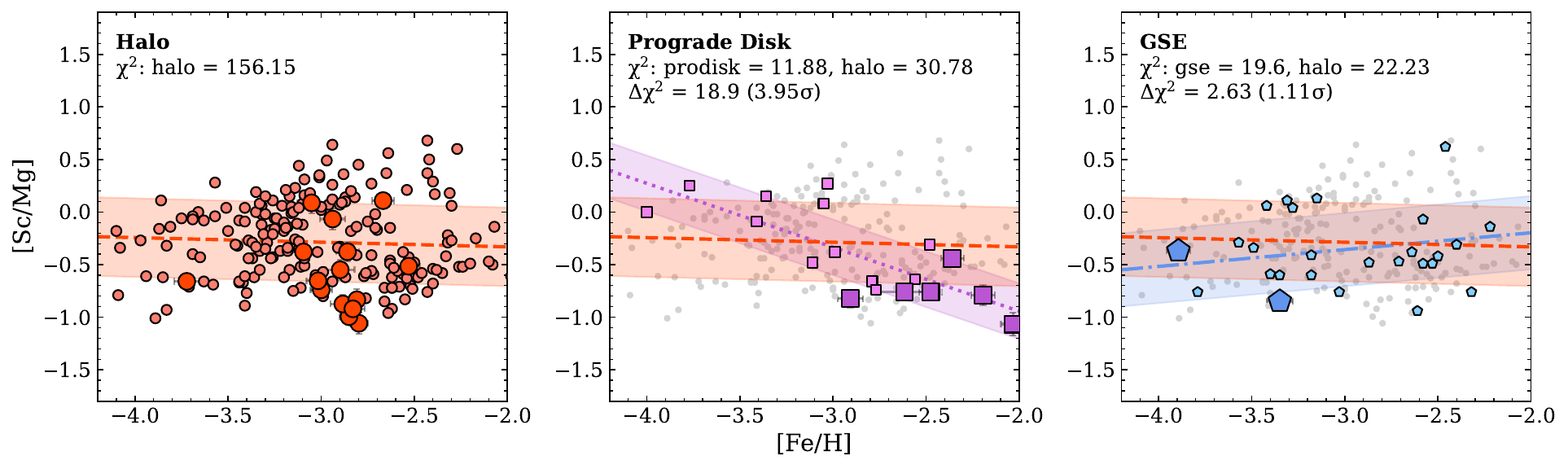}
    \caption{The $\Delta \chi^2$ results for $\XFe{Mg}$ (top row), $\XFe{Sc}$ (middle row) and $\XY{Sc}{Mg}$ (bottom row) as a function of $\FeH$ across the prograde disk (purple squares; middle figure) and GSE (blue pentagons; right figure) regions. Note that for each row, the halo region (red circles; left figure) was taken as the null hypothesis, hence no calculations were made on it. The grey points in both the prograde disk and GSE figures represent the halo points from the first figure in each row. For all regions, we show the linear fit derived from the halo dataset, as shown by the orange dashed-line. The $\chi^2$ from this in all three regions are printed in each figure. For both the prograde disk and GSE regions, we show their individual fits, represented by the purple dotted and blue dot-dashed lines in their respective figures. Their $\chi^2$ values are printed alongside the halo $\chi^2$. For these two regions, the $\Delta \chi^2$ results were calculated (original fit $\chi^2$ minus halo fit $\chi^2$), printed beneath the individual $\chi^2$ values.} 
    \label{fig:scmg plot}
\end{figure*}

Two elements show abundance trends that disagree at the $3\sigma$ level between the prograde disk and halo: Mg with $\Delta \chi^2 = 14.22$ at $3.3\sigma$ confidence (first row in Fig.~\ref{fig:scmg plot}), and Sc with $\Delta \chi^2 = 29.50$ at $5.07\sigma$ confidence (second row in Fig.~\ref{fig:scmg plot}). To isolate the relative production of these elements, we evaluated the abundance ratio $\XY{Sc}{Mg}$, as shown in Fig.~\ref{fig:scmg plot} as a function of $\FeH$. We see that the linear fit to the prograde disk data (purple dotted line with $\XY{Sc}{Mg}_{\rm prodisk} = -0.61\FeH - 2.61$), produces a better $\chi^2$ value at $\chi^2_{\rm prodisk} = 11.88$, than the fit from the halo-derived line at $\chi^2_{\rm halo} = 30.78$ (with $\XY{Sc}{Mg}_{\rm halo} = -0.04\FeH - 0.42$), providing a high $\Delta \chi^2 = 18.9$ value. At $3.95\sigma$, this shows that the prograde disk fit is significantly different from the halo in $\XY{Sc}{Mg}$-$\FeH$ abundance space. This is not seen in the GSE region (with $\XY{Sc}{Mg}_{\rm GSE} = +0.16\FeH + 0.12$), showing a significantly lower $\Delta \chi^2 = 2.63$ value at $1.11\sigma$, suggesting similar $\XY{Sc}{Mg}$ enrichment as for the halo. We note that while the prograde disk trend is predominantly driven by our higher-metallicity X-Shooter targets, the absence of a similar offset in our halo sample confirms this feature is astrophysical rather than systematic.

\begin{figure}
    \centering
    \includegraphics[width=1\linewidth]{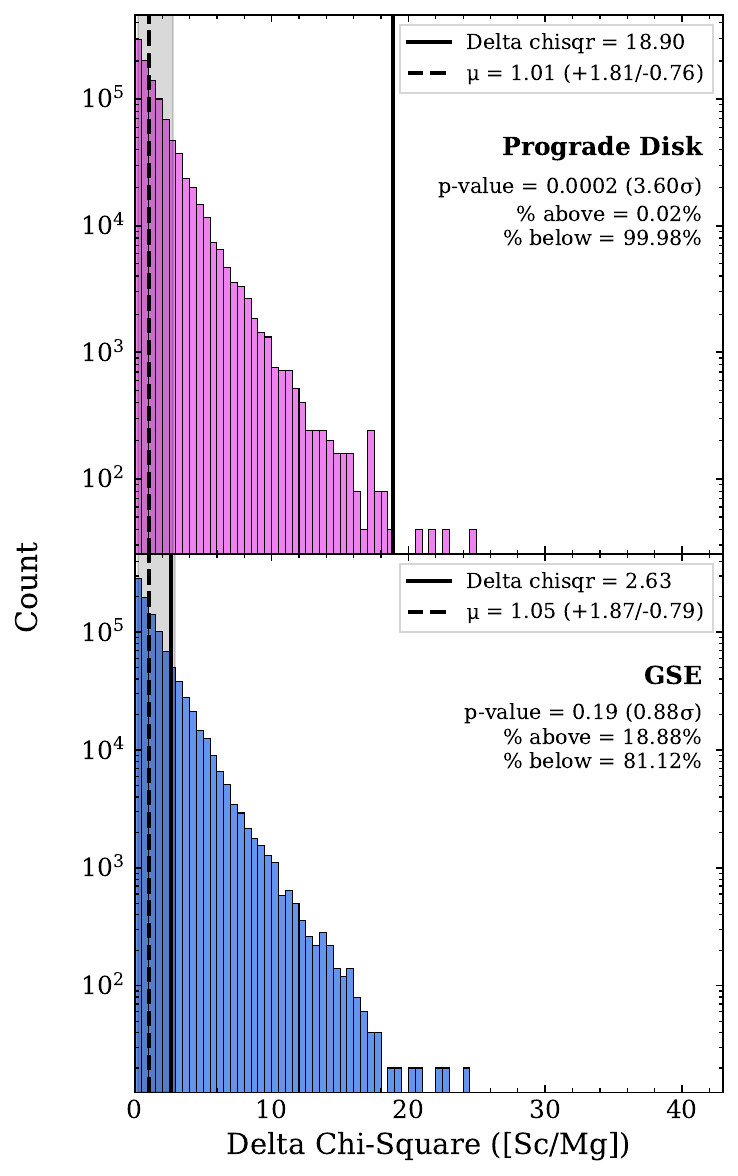}
    \caption{Distribution of $\Delta \chi^2$ values for $\XY{Sc}{Mg}$ across the prograde disk (top panel) and GSE (bottom panel) regions for one million iterations. The median of the distribution is shown by the vertical black-dashed line, with the $1\sigma$ range shown by the black shaded region. The original $\Delta \chi^2$ of the regions from Fig.~\ref{fig:scmg plot} is shown by the vertical black line. For each iteration, halo stars were randomly selected based on the number of prograde disk/GSE stars in bins of $+0.5$\,dex (with $\pm0.1$\,dex jitter) across our metallicity range. For the randomly-selected stars, a linear fit was derived, then using the $\chi^2$ value, it was used in the $\Delta \chi^2$ calculation alongside the original halo fit's $\chi^2$ to the resampled data. The percentage of iterations having $\Delta \chi^2$ values above and below the original $\Delta \chi^2$, alongside the p-value, is printed on the plot.} 
    \label{fig:scmg resampling plot}
\end{figure}

To quantify the statistical likelihood that the trends seen in Fig.~\ref{fig:scmg plot} could arise by chance, we performed a Monte Carlo resampling test on both the GSE and prograde disk populations. For this, we randomly selected halo stars based on the number of prograde disk/GSE stars in bins of $+0.5$\,dex (with $\pm0.1$\,dex jitter) across our metallicity range. Once drawn, the linear fit was re-derived with the resampled halo data, followed up by the $\Delta \chi^2$ calculation with the original halo fit to the data. This was repeated one million times, creating the distribution seen in Fig.~\ref{fig:scmg resampling plot}. 

For the prograde disk, we see that from the original $\Delta \chi^2_{\rm prodisk} = 18.9$ value, only $0.02$\% of the iterations have a greater $\Delta \chi^2$ value. This contrasts with the GSE, which at $\Delta \chi^2_{\rm GSE} = 2.63$, has $18.62$\% of the iterations with a larger $\Delta \chi^2$ value. 

In our work, we adopted the MWPotential2014 Galactic potential due to consistency with our previous work. However, to confirm that the $\XY{Sc}{Mg}$ trend in the prograde disk was not dependent upon choice of Galactic potential, we recalculated the stellar orbits for the combined X-Shooter and Literature Comparison sample using the \citet{mcmillan_mass_2017} Galactic potential with the \citet{binney_actions_2012} St\"{a}ckel approximation. This potential assumes a heavier Milky Way mass ($\approx 1.3 \times 10^{12}$\,M$_{\odot}$) than MWPotential2014 (which assumes a lighter mass at $\approx 0.8 \times 10^{12}$\,M$_{\odot}$). Additionally, the \citet{mcmillan_mass_2017} potential assumes four disk components (thin disk, thick disk, HI and H2 gas), rather than the single merged disk assumed by MWPotential2014. 

With the orbits re-derived, we re-assigned our stars to the four kinematic groupings. For the prograde disk, $19$ prograde disk stars were classified ($17$ Literature Comparison, $2$ X-Shooter), with $16$ of these stars ($14$ Literature Comparison, $2$ X-Shooter) already belonging to the prograde disk in the MWPotential2014 analysis. The remaining $3$ stars become halo stars in the \citet{mcmillan_mass_2017} potential. For the GSE, $30$ stars were classified ($28$ Literature Comparison, $2$ X-Shooter), with $22$ of these stars ($20$ Literature Comparison, $2$ X-Shooter) already belonging to the GSE in the MWPotential2014 analysis. The remaining $8$ stars become halo stars in the \citet{mcmillan_mass_2017} potential. For completeness, we found that with the \citet{mcmillan_mass_2017} potential, $6$ stars were classified as part of the retrograde disk (all Literature Comparison), with $4$ of them already being classified in the retrograde disk in the MWPotential2014 analysis. 

The steps described above to generate our $\XY{Sc}{Mg}$ result was repeated here with the \citet{mcmillan_mass_2017} analysis. For the prograde disk, we found that the slope of the fit with the classification from the \citet{mcmillan_mass_2017} potential, $-0.60$ ($\Delta \XY{Sc}{Mg}/\Delta \FeH$), is very consistent with the slope of the relation ($-0.61$\,dex per dex) found with the MWPotential2014 classification. Despite the $y$-intercepts between the two relations being inconsistent ($c_{\rm mcmillan17} = -2.14$ and $c_{\rm mwpotential14} = -2.61$), the resulting $\Delta \chi^2$ result, with the \citet{mcmillan_mass_2017}, is still above $3\sigma$ statistical confidence at $\Delta \chi^2_{\rm mcmillan17} = 13.87$ with $3.30\sigma$. This demonstrates that the trend seen in Fig.~\ref{fig:scmg plot} is statistically significant, and not dependent on the choice of Galactic potential. We discuss these findings in the next Section.

\section{Discussion}
\label{sec:discussion}
We see in the top panel of Fig.~\ref{fig:scmg plot} that at $\FeH < -3.0$, the halo has higher $\XFe{Mg}$ than it does at $\FeH > -3.0$. This is the opposite for the prograde disk, where we see lower $\XFe{Mg}$ at lower metallicities, before it becomes more abundant at higher metallicities. Typically, at low metallicities, Mg is produced through hydrostatic burning in massive stars, and then released through Type II supernovae \citep{timmes_galactic_1995, kobayashi_galactic_2006, kobayashi_origin_2020}. Contribution of massive stars is seen by the plateau in $\alpha$-abundances with increasing metallicity that is terminated by the onset of Type Ia SN leading to a 'knee' in the $\XFe{Mg}$, $\FeH$ relation \citep[e.g.][]{tinsley_evolution_1980, mcwilliam_spectroscopic_I_1995, mcwilliam_spectroscopic_II_1995, matteucci_chemical_2001}. For the prograde disk, the lower $\XFe{Mg}$ abundances at $\FeH < -3.0$ likely indicates that there was reduced Mg enrichment in the progenitor prograde disk, caused by e.g. lower massive star formation rates at early times. At $\FeH > -3.0$, the increasing $\XFe{Mg}$ in the prograde disk could be the result of significant gas in-fall, increasing the star formation rate and thus increasing Mg enrichment. Interestingly, the trend for Mg is not seen in the other $\alpha$-elements studied in this work, which all show consistency between the three kinematic regions (see Appendix \ref{appendix: delta chi2}).

For Sc, we see in the middle panel of Fig.~\ref{fig:scmg plot} that at $\FeH > -3.0$, the prograde disk has lower $\XFe{Sc}$ than what we see in the halo or the GSE. At $\FeH < -3.0$, we see that the differences in $\XFe{Sc}$ between the kinematic populations is indistinguishable. The nucleosynthetic origin of Sc is not well understood, with it always being under produced in detailed supernova simulations \citep[e.g.][]{woosley_evolution_1995, kobayashi_galactic_2006}. Generally though, Sc is produced in explosive burning in massive stars and core collapse supernovae \citep[e.g.][]{kobayashi_origin_2020}. How this relates to what we see in the prograde disk at higher metallicities is yet to be understood.

The results we see for $\XY{Sc}{Mg}$ in our prograde disk sample is independent of our kinematic classifications under different Milky Way potential models. It is also unlikely that our $\XY{Sc}{Mg}$ prograde disk result is caused by systematics in our spectroscopic analysis, since we do not see it in the other kinematic populations. At this stage, without more information on the gas distribution at these early times, alongside a better understanding on the origin of Sc, we can not discern the implications of our results in the context of the formation processes for the metal-weak prograde disk.

\section{Summary and Conclusion}
\label{sec:conclusion}
Starting from our 2dF+AAOmega survey of metal-poor disk stars \citep{lowe_rise_2025}, in this work we have extended our X-Shooter follow up of the most metal-poor candidates by including seven additional stars (P115 sample) to the $16$ already discussed in \citet[][P113 sample]{lowe_detailed_2026}. All these stars were reduced and analysed in the same way, deriving metallicities and measuring chemical abundances for $16$ elements (C (CH), N (NH), \ion{Na}{I}, \ion{Mg}{I}, \ion{Al}{I}, \ion{Si}{I}, \ion{Ca}{II}, \ion{Sc}{II}, \ion{Ti}{II}, \ion{Cr}{I}, \ion{Mn}{I}, \ion{Co}{I}, \ion{Ni}{I}, \ion{Sr}{II}, \ion{Ba}{II} and \ion{Eu}{II}). The chemical abundances for both the P115 and P113 stars are in excellent agreement with the Comparison Literature sample \citep[Fig.~\ref{fig:xfe abunds};][]{yong_most_2013, jacobson_high-resolution_2015, marino_keck_2019, yong_high-resolution_2021}.

Among our P115 sample, we have one C-poor star (Fig.~\ref{fig:c enhancement}), one probable and two possible NEMP stars (Fig.~\ref{fig:n enhancemenet}), one $\alpha$-enhanced star (Fig.~\ref{fig:alpha enhancemenet}), and no stars with negative $\XY{Sr}{Ba}$ values (Fig.~\ref{fig:srba plot}). In particular, the abundance trends for the sample (Fig.~\ref{fig:xshooter abund trend}) reveal the peculiar halo star ra\_0752-5047\_s47 with $\FeH = -3.05 \pm 0.08$, is strongly enhanced in Ti ($\XFe{Ti} = +0.73 \pm 0.04$, making it $\alpha$-enhanced at $\aFe = +0.54 \pm 0.02$). It also has a moderate enhancement in Sc ($\XFe{Sc} = +0.57 \pm 0.07$) and is C-poor ($\XFe{C}_{\rm corrected} = -0.31 \pm 0.07$, with low evolutionary correction of $+0.01$\,dex). Interestingly, it is also an r-I star ($\XFe{Eu} = +0.73 \pm 0.10$ and $\XY{Ba}{Eu} = -1.29 \pm 0.11$). A massive jet-induced hypernovae \citep{tominaga_aspherical_2009} may feasibly be the progenitor to create these abundance patterns, alongside a separate r-process event. Future high-resolution spectroscopy to measure Zn (predicted to be overproduced) is needed to confirm this, as an overabundance is characteristic of large explosion energy supernovae (hypernovae) \citep{nomoto_nucleosynthesis_2013}. We also find star ra\_1639-2632\_s419 with $\FeH = -2.03 \pm 0.06$, having a low N upper-limit of $\XFe{N} < -0.90$. However, it lacks the other peculiarities of the analogous N-poor star in the P113 sample.

From the combined X-Shooter (23 stars) and Literature Comparison (315 stars, all with \textit{Gaia} DR3 entries) samples, we derived orbits using the MWPotential2014 Galactic potential, separating the stars out into prograde disk ($6.5$\%), retrograde disk ($1.2$\%), GSE ($10.7$\%) and halo ($81.7$\%) kinematic classifications (Fig.~\ref{fig:circle plot}). For each classification and $\XFe{X}$ combination, linear fits were computed across metallicities $-4.2 \leq \FeH \leq -1.9$. Excluding the retrograde disk region (due to small sample size), we ran a $\Delta \chi^2$ test on the prograde disk and the GSE fits to see if they were chemically-distinct from the halo fit (Table~\ref{tab:chemodynamic table} and Appendix \ref{appendix: delta chi2}). For the GSE region, we showed that there were no $\XFe{X}$-$\FeH$ evolutionary trends that were statistically different from the halo. For the prograde disk, we revealed that both Mg and Sc were statistically significant compared with the halo ($3\sigma$ and $5\sigma$ respectively). To isolate elemental production, we took $\XY{Sc}{Mg}$ as a function of $\FeH$ (Fig.~\ref{fig:scmg plot}), again showing that the prograde disk is chemically distinct from the halo, with a statistical significance of $3.95\sigma$. Unlike the halo or the GSE, which both have relatively flat slopes at $m_{\rm halo} = -0.04x$ and $m_{\rm GSE} = 0.16x$ respectively, the prograde disk has a strongly negative $\XY{Sc}{Mg}$ slope at $m_{\rm prodisk} = -0.61\FeH$.

We tested the robustness of this trend by running two tests. The first was a resampling Monte Carlo test with one million iterations (Fig.~\ref{fig:scmg resampling plot}), revealing that only $0.02$\% of the iterations had a statistical significance greater than $3.95\sigma$. The second test was to re-run the orbital derivations for the combined sample with the \citet{mcmillan_mass_2017} Galactic potential, showing that the prograde disk trend in $\XY{Sc}{Mg}$ was not potential-dependent, reproducing the behaviour from the MWPotential2014 model with a statistical significance of $3.30\sigma$. 

The unexpected $\XY{Sc}{Mg}$ trend for the prograde disk is driven by the relatively low $\XFe{Mg}$ at $\FeH < -3.0$, and the relatively low $\XFe{Sc}$ at $\FeH > -3.0$, a trend absent from both the GSE and halo populations. We suggest that this is in part due to delayed Mg enrichment within the progenitor prograde disk, possibly driven by lower massive star formation rates, before experiencing subsequent Mg enrichment from an unknown source like a gas accretion event. However, future advancements in the origin of Sc are needed to confirm this.

\section*{Acknowledgements}
This paper includes data gathered with the $8$\,m VLT located at Cerro Paranal, Chile, and is based on observations collected at the European Southern Observatory under ESO programs 113.26N5.001 and 115.28C7.001. 

This work was supported by computational resources provided by the Australian Government through the National Computational Infrastructure (NCI) under the National Computational Merit Allocation Scheme (project y89). T.N. acknowledges support from the Knut and Alice Wallenberg Foundation.

We thank the anonymous referee for their insightful comments and suggestions. Their feedback has helped improve the quality and strength of this manuscript.


\section*{Data Availability}
The data used in this study are available in the ESO archive (\url{https://archive.eso.org/eso/eso_archive_main.html}) under programme IDs 113.26N5.001 and 115.28C7.001. Our co-added spectra are available upon request.



\bibliographystyle{mnras}
\bibliography{references} 




\section*{Supporting Information}
Tables in the manuscript (Tables 1 -- 10) and supplementary material (Tables S1 -- S4) are available at \textit{MNRAS} online. For Tables 1 -- 3, Tables 5 -- 8 and Tables S2 -- S4, we also provide entries for the P113 stars from \citetalias{lowe_detailed_2026}.

\begin{description}
    \item[\textbf{Table 1}] Program stars and observational parameters.
    \item[\textbf{Table 2}] Observing details.
    \item[\textbf{Table 3}] Stellar parameters.
    \item[\textbf{Table 4}] Median abundance errors.
    \item[\textbf{Table 5}] Chemical abundances.
    \item[\textbf{Table 6}] $\XY{C}{N}$ abundances.
    \item[\textbf{Table 7}] $\rm{[\alpha/Fe]}$ abundances.
    \item[\textbf{Table 8}] $\XY{Sr}{Ba}$ and $\XY{Ba}{H}$ abundances.
    \item[\textbf{Table 9}] Kinematic classifications.
    \item[\textbf{Table 10}] $\Delta \chi^2$ values.
\end{description}

\begin{description}
    \item[\textbf{Table S1}] Fitting windows.
    \item[\textbf{Table S2}] Fitting windows for Na.
    \item[\textbf{Table S3}] NLTE corrections and continuum offsets for \ion{Ca}{II}.
    \item[\textbf{Table S4}] CH evolutionary corrections.
\end{description}

Please note: Oxford University Press is not responsible for the content or functionality of any supporting materials supplied by the authors. Any queries (other than missing material) should be directed to the corresponding author for the article.
\appendix
\section{CH Fits}
\label{appendix: ch fits}
\begin{figure*}  
    \centering
     \begin{subfigure}{0.49\textwidth}
        \centering
        \includegraphics[width=0.85\linewidth]{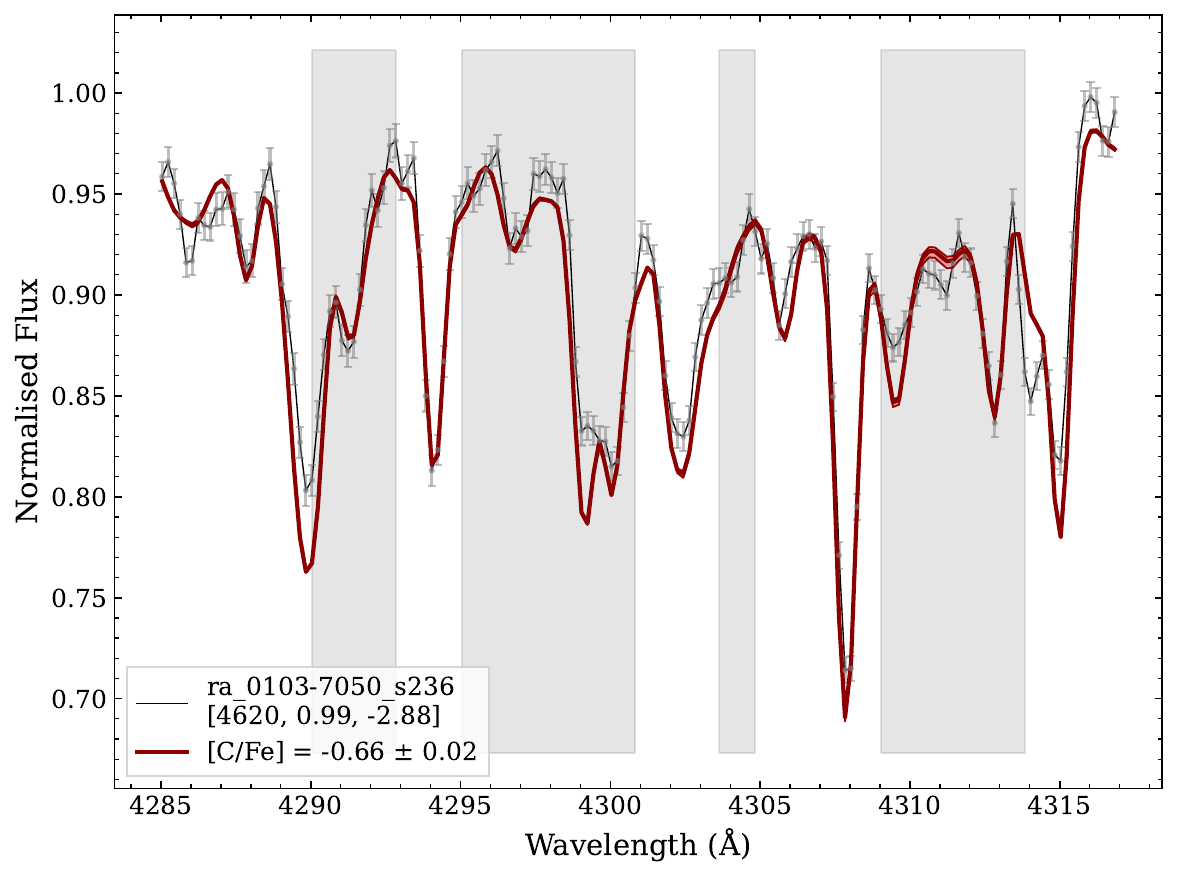}
    \end{subfigure}
    \hfill
    \begin{subfigure}{0.49\textwidth}
        \centering
        \includegraphics[width=0.85\linewidth]{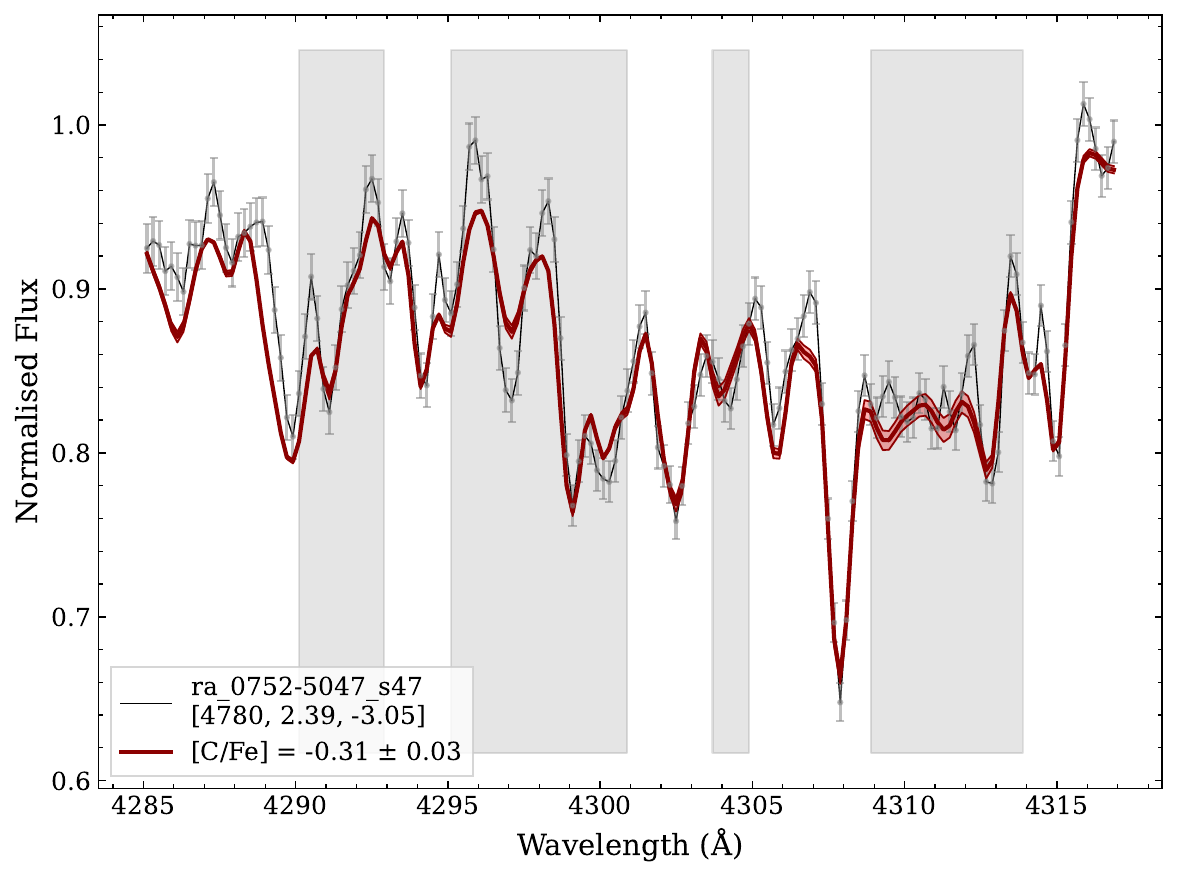}
    \end{subfigure}

    \begin{subfigure}{0.49\textwidth}
        \centering
        \includegraphics[width=0.85\linewidth]{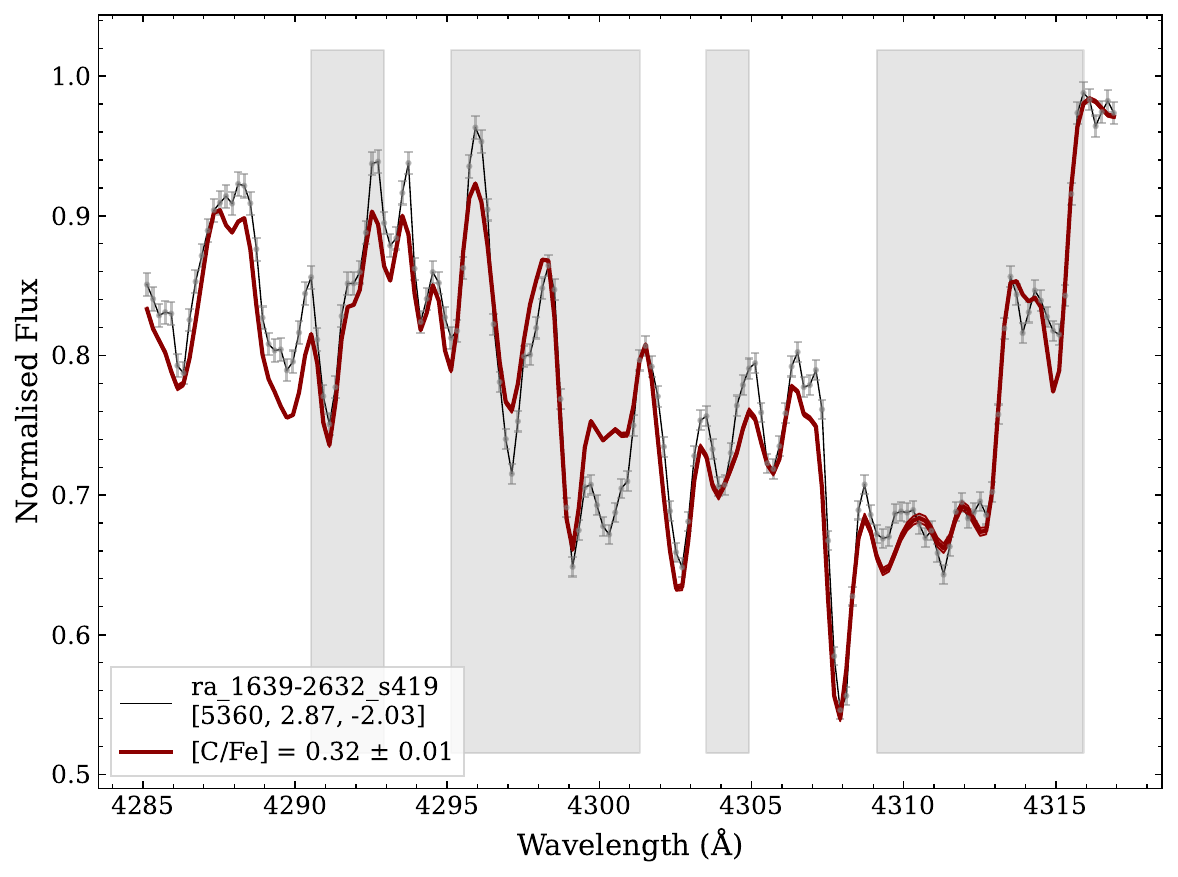}
    \end{subfigure}
    \hfill
    \begin{subfigure}{0.49\textwidth}
        \centering
        \includegraphics[width=0.85\linewidth]{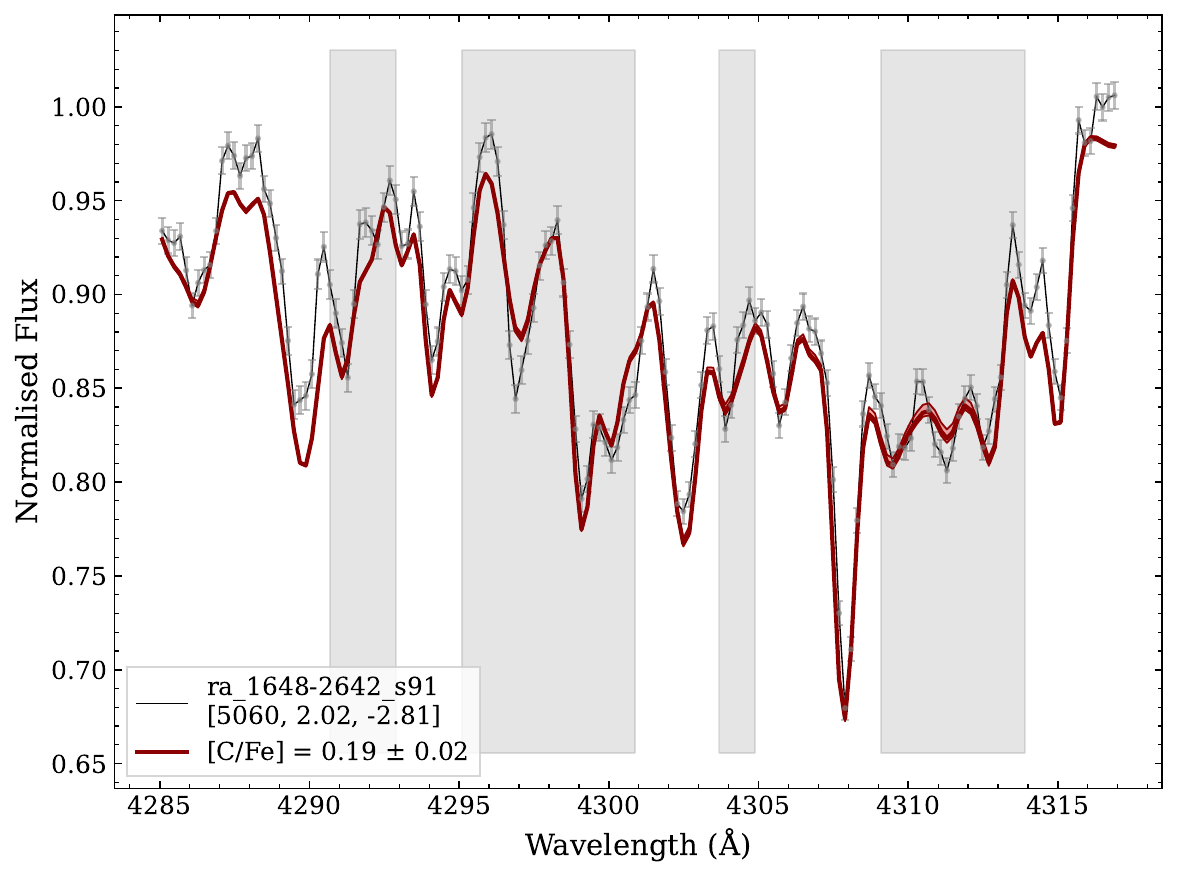}
    \end{subfigure}

    \begin{subfigure}{0.49\textwidth}
        \centering
        \includegraphics[width=0.85\linewidth]{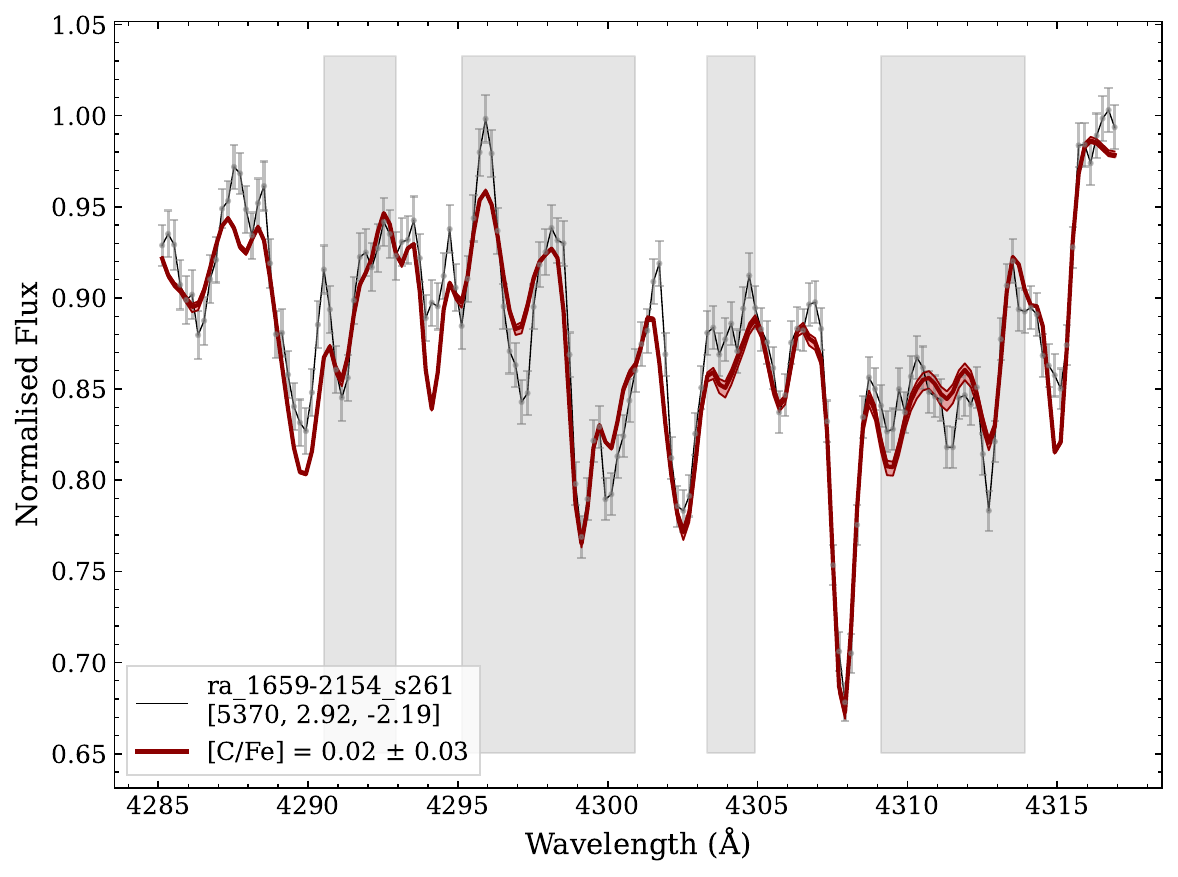}
    \end{subfigure}
    \hfill
    \begin{subfigure}{0.49\textwidth}
        \centering
        \includegraphics[width=0.85\linewidth]{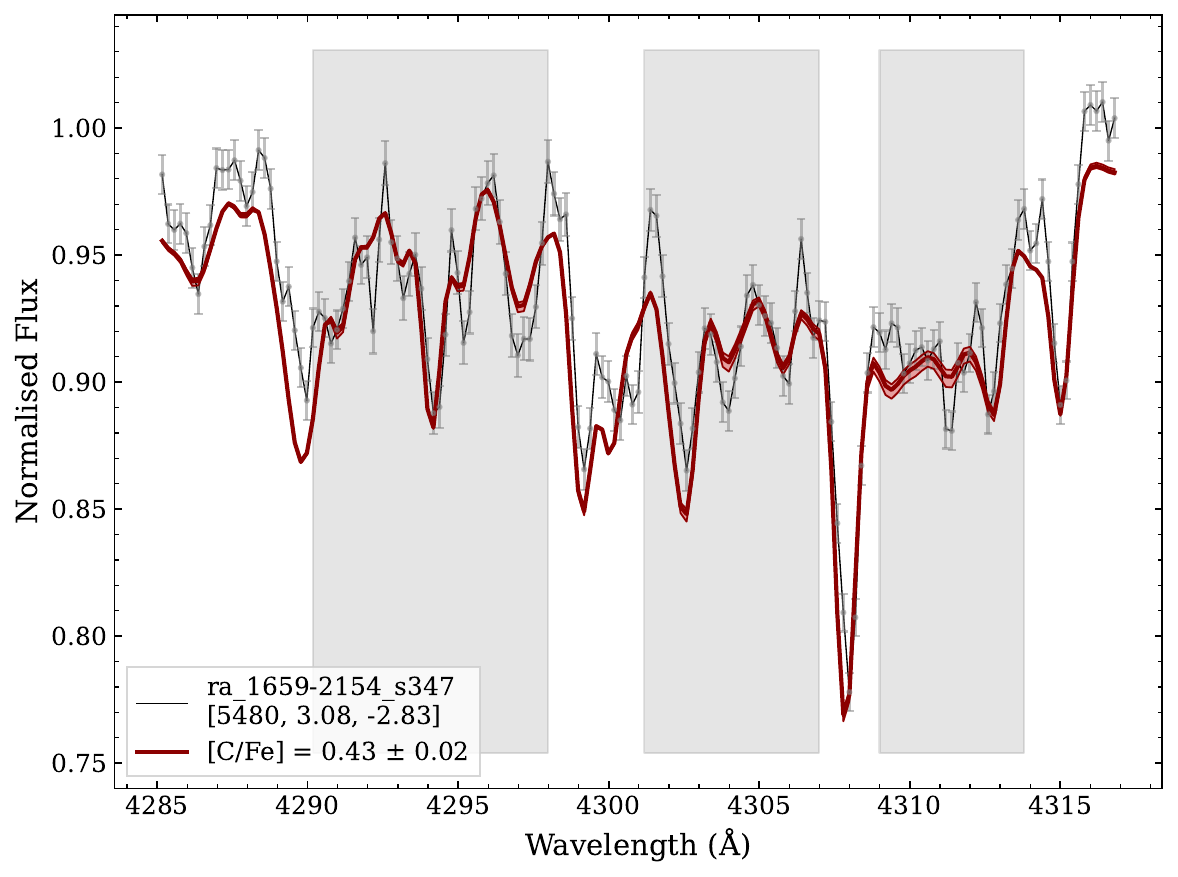}
    \end{subfigure}
    
    \begin{subfigure}{0.49\textwidth}
        \centering
        \includegraphics[width=0.85\linewidth]{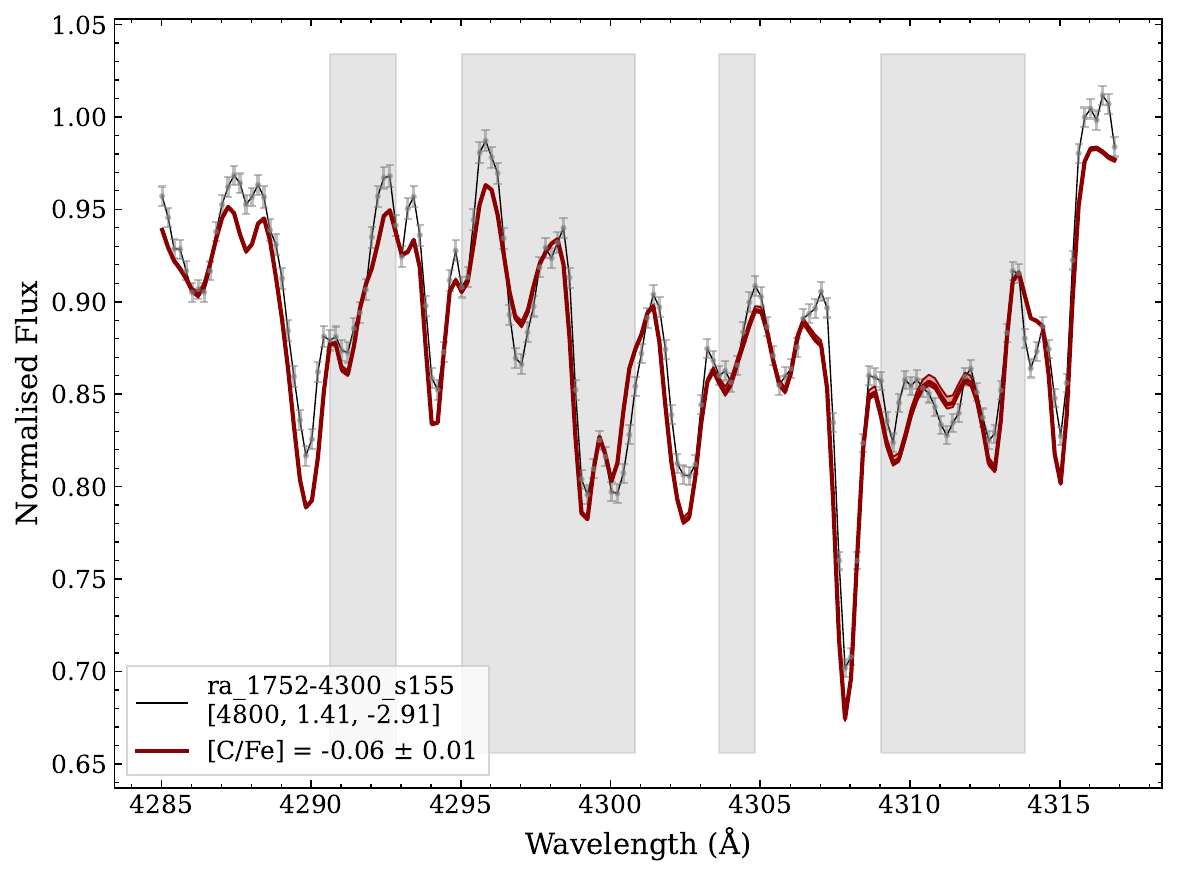}
    \end{subfigure}
    \caption{CH fits for the seven P115 stars across the wavelength region $4285 \leq \lambda \leq 4317$\,\AA{}. The observed data is in black, with the red line representing the best-fitted $\XFe{C}$ value (alongside its fitting error; $\XFe{C}$ value not corrected for evolutionary effects). The statistical error shown by the red shaded region. Stellar parameters $\Teff$, $\logg$ and $\FeH$ are found in the legend for each star. Grey shaded regions refer to regions used for $\chi^2$ calculations. Several prominent atomic lines are present, including \ion{Fe}{I} at $4294.125$ and $4307.901$\,\AA{}, alongside \ion{Ca}{I} at $4302.528$\,\AA{}.}
    \label{fig:cfe fits}
\end{figure*}

We show the best-fitting $\XFe{C}$ spectrum to the observed P115 stellar CH regions across wavelengths $4285 \leq \lambda \leq 4317$\,\AA{} in Fig.~\ref{fig:cfe fits}. From the P115 sample, all seven stars have CH detections, with no non-detections reported.

\section{NH Fits}
\label{appendix: nh fits}
\begin{figure*}  
    \centering
     \begin{subfigure}{0.49\textwidth}
        \centering
        \includegraphics[width=0.85\linewidth]{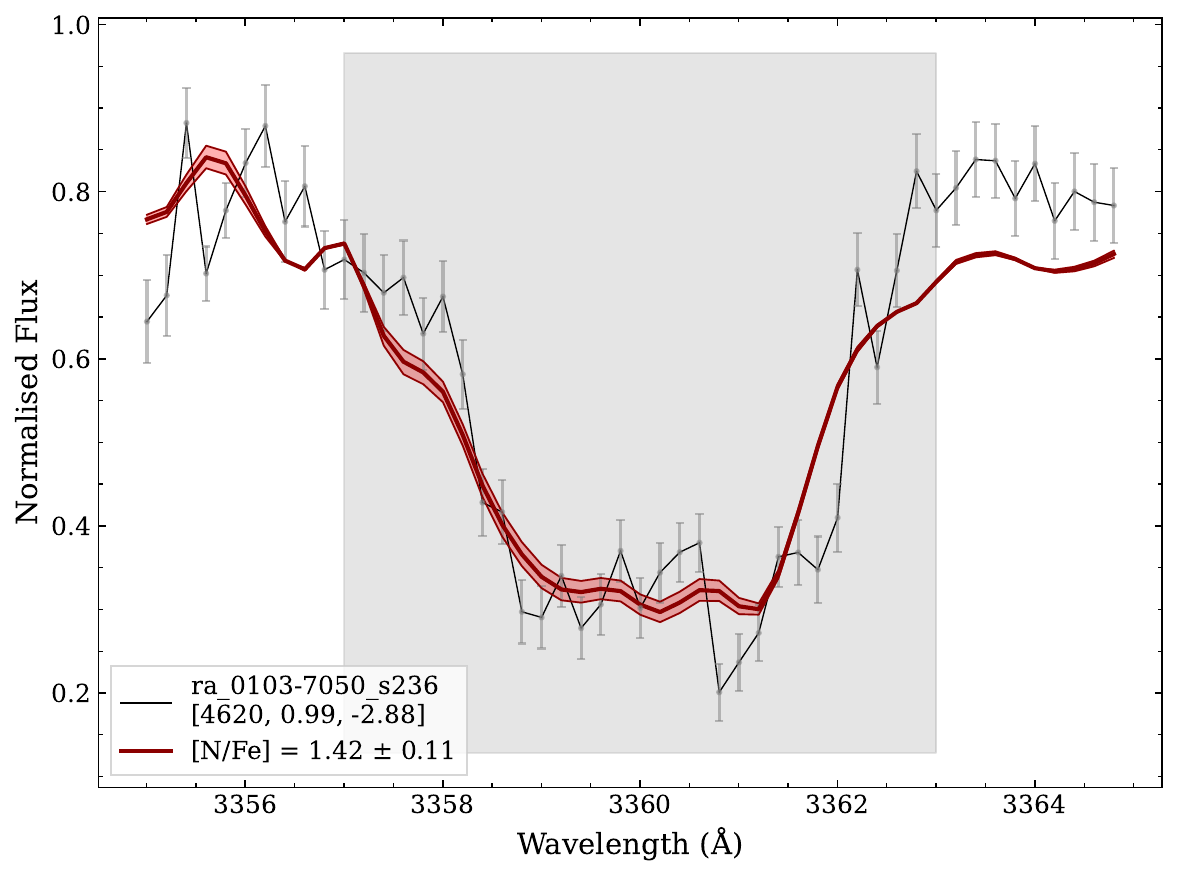}
    \end{subfigure}
    \hfill
    \begin{subfigure}{0.49\textwidth}
        \centering
        \includegraphics[width=0.85\linewidth]{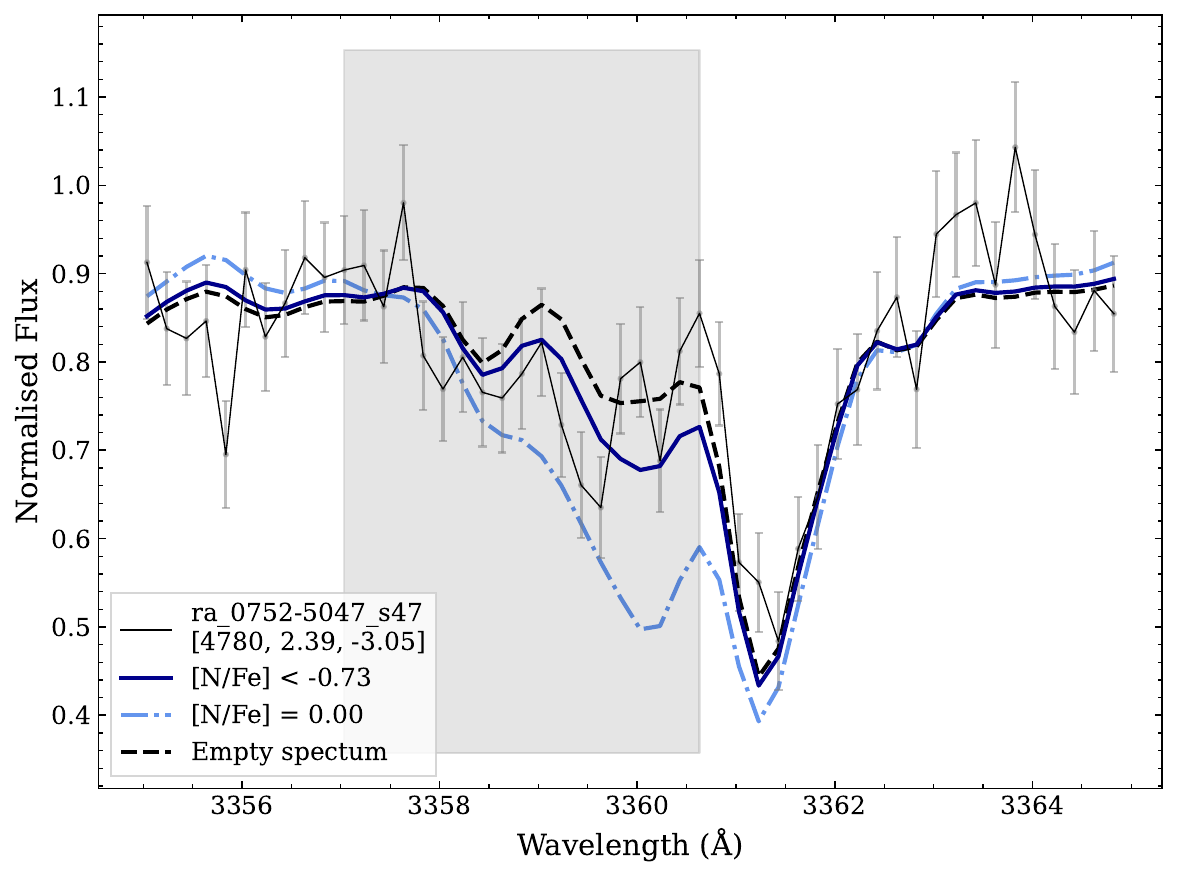}
    \end{subfigure}

    \begin{subfigure}{0.49\textwidth}
        \centering
        \includegraphics[width=0.85\linewidth]{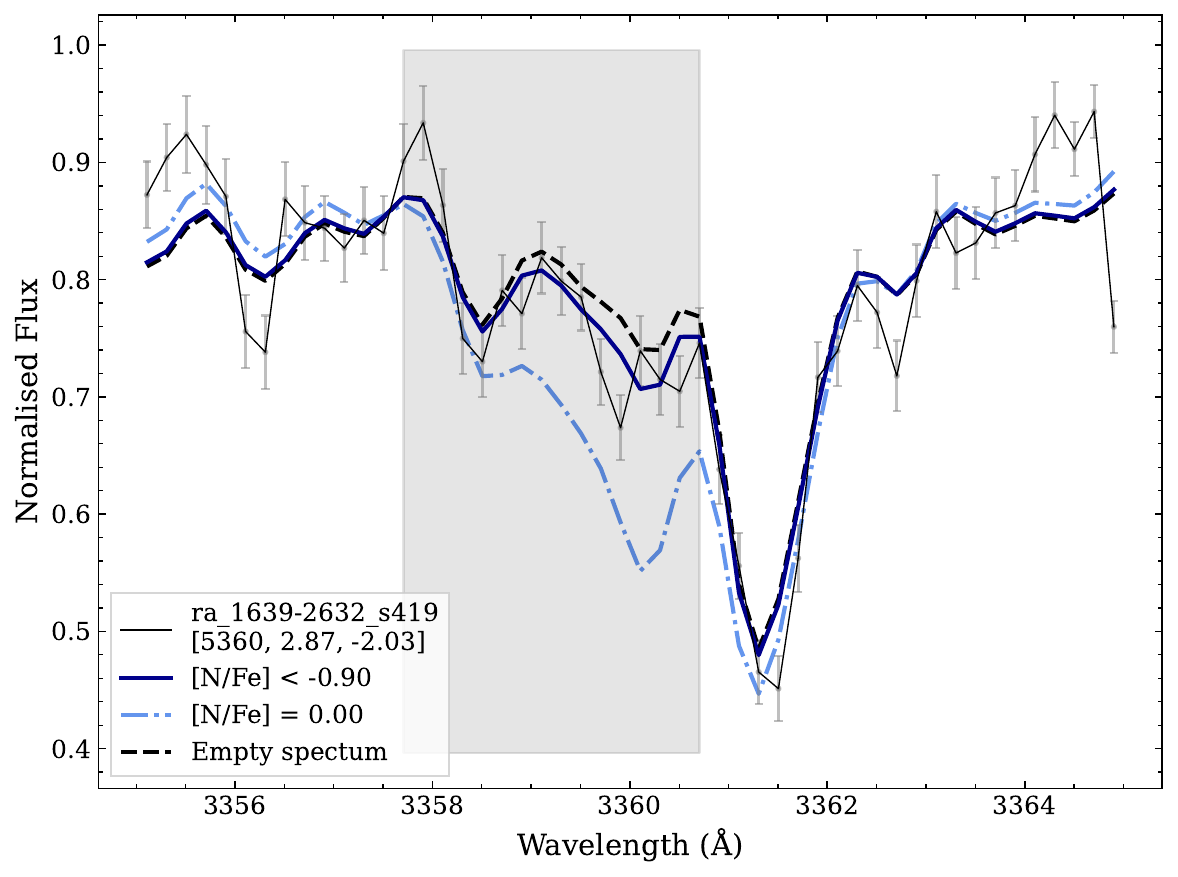}
    \end{subfigure}
    \hfill
    \begin{subfigure}{0.49\textwidth}
        \centering
        \includegraphics[width=0.85\linewidth]{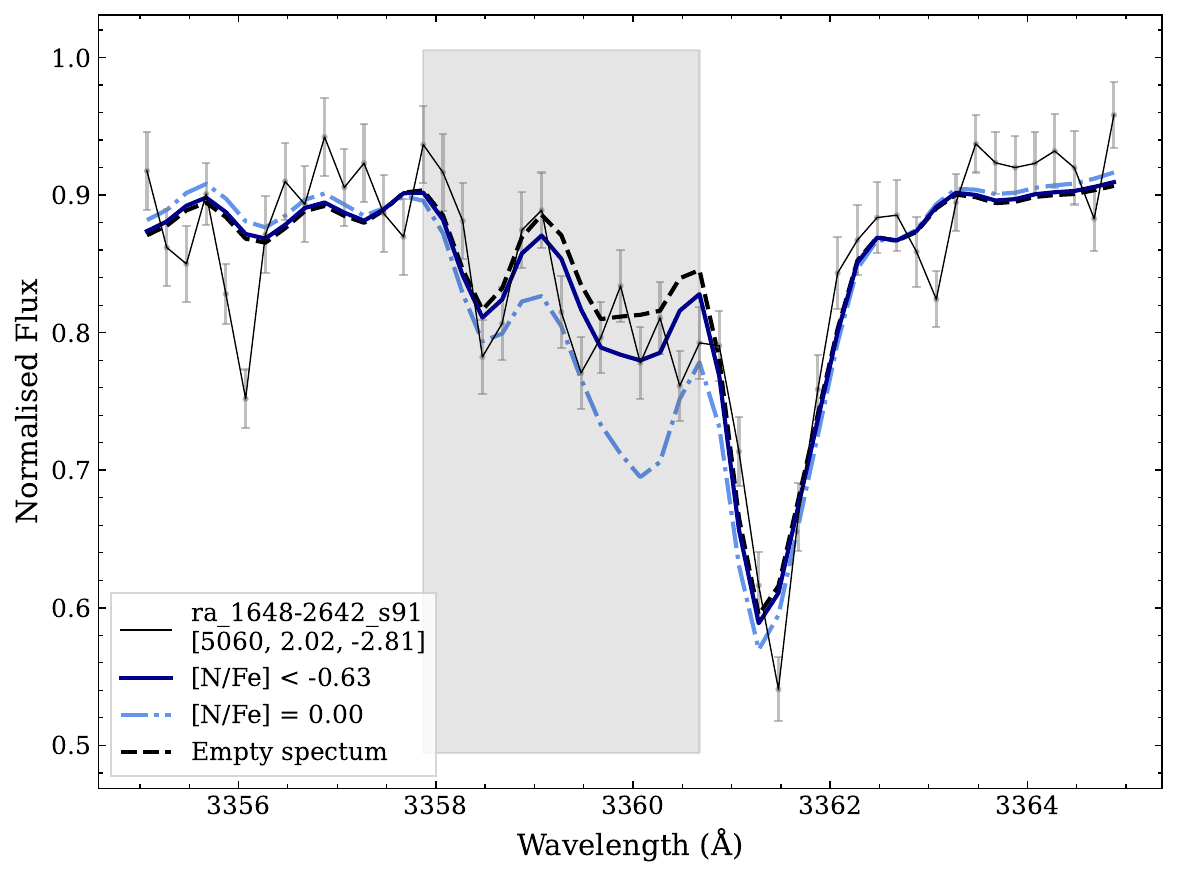}
    \end{subfigure}

    \begin{subfigure}{0.49\textwidth}
        \centering
        \includegraphics[width=0.85\linewidth]{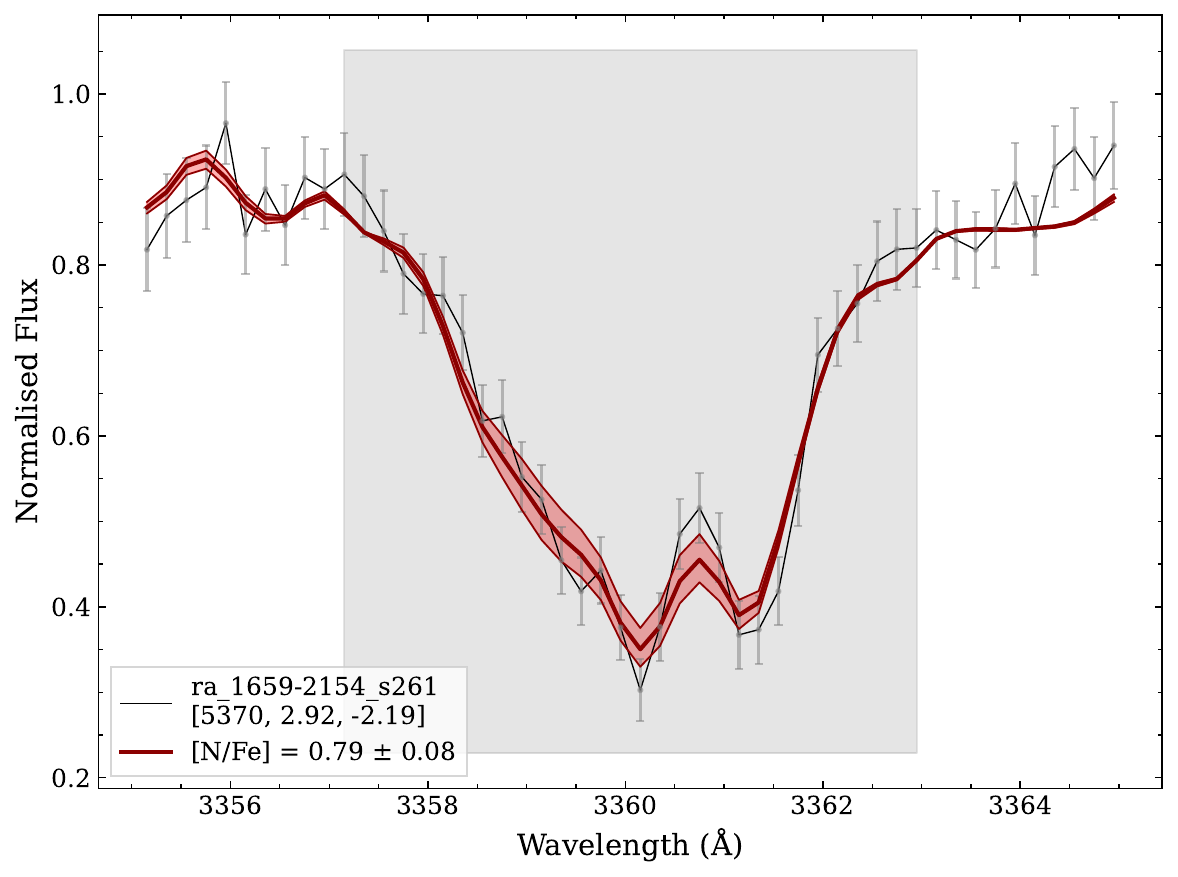}
    \end{subfigure}
    \hfill
    \begin{subfigure}{0.49\textwidth}
        \centering
        \includegraphics[width=0.85\linewidth]{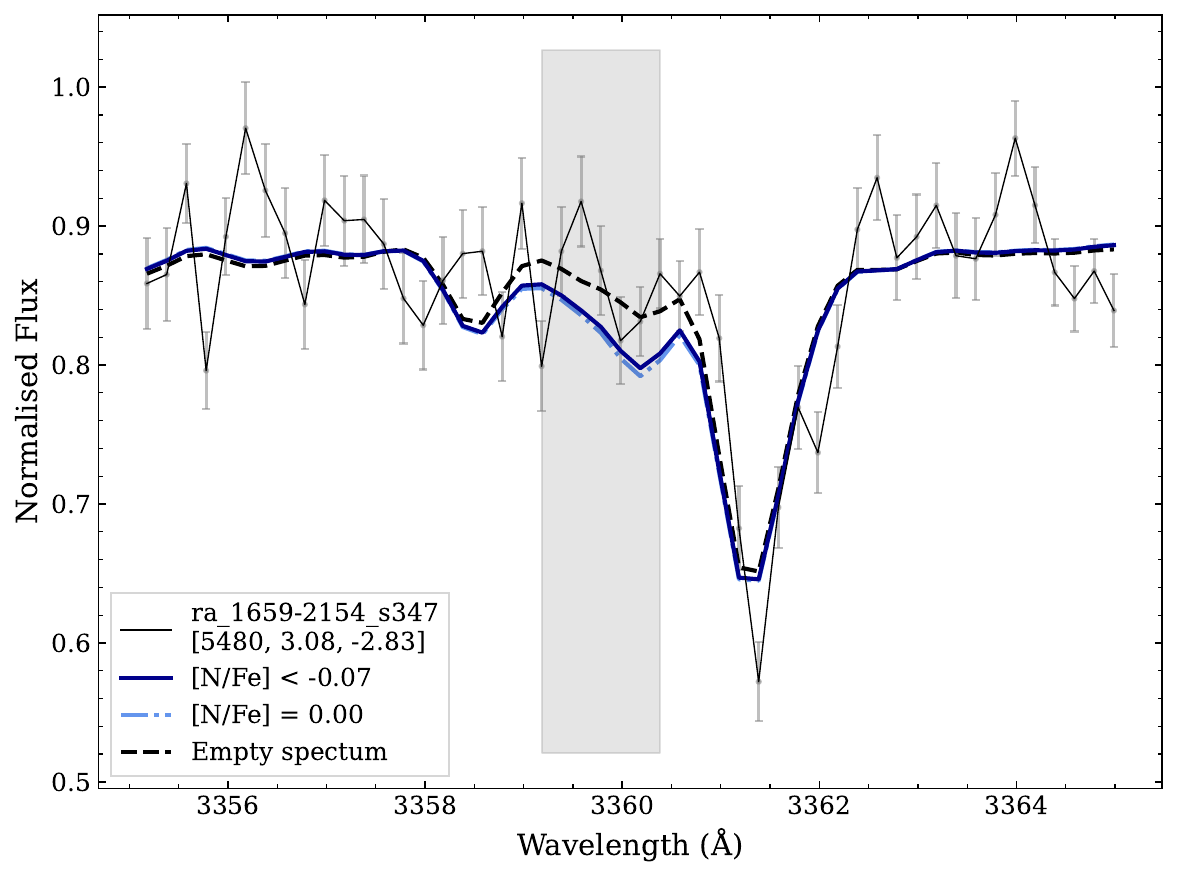}
    \end{subfigure}
    
    \begin{subfigure}{0.49\textwidth}
        \centering
        \includegraphics[width=0.85\linewidth]{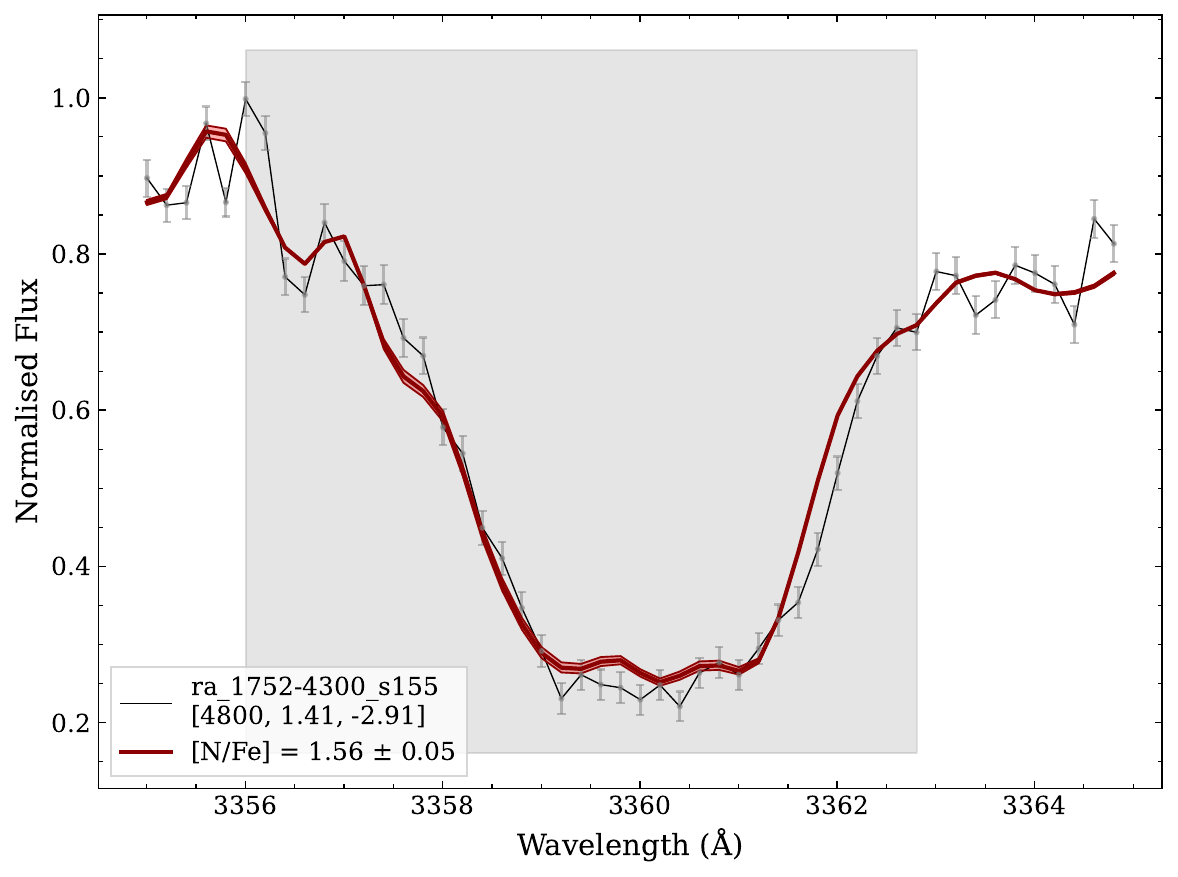}
    \end{subfigure}
    \caption{NH fits for the seven P115 stars across the wavelength region $3355 \leq \lambda \leq 3365$\,\AA{}. Alongside the formatting used in Fig.~\ref{fig:cfe fits}, stars with non-detections have their upper limit values fitted instead, shown by the blue line. For non-detection cases, two reference synthetic spectra are plotted: one with $\XFe{N} = 0.0$, represented by the dashed light blue line, and the other with $\XFe{N} = -3.0$, represented by the dashed black line as the `empty' spectrum. Across the wavelength region, atomic lines \ion{Cr}{I} at $3358.491$\,\AA{} and \ion{Ti}{II} at $3361.212$\,\AA{} are present.}
    \label{fig:nfe fits}
\end{figure*}

We show the best-fitting $\XFe{N}$ spectrum to the observed P115 stellar NH regions across wavelengths $3355 \leq \lambda \leq 3365$\,\AA{} in Fig.~\ref{fig:nfe fits}. From the sample, four stars have non-detections, with the remaining three having detections in NH. The stars with non-detections are represented by their upper limits instead.

\section{$\Delta \chi^2$ results}
\label{appendix: delta chi2}
\begin{figure*}  
    \centering
     \begin{subfigure}{\textwidth}
        \centering
        \includegraphics[width=1\linewidth]{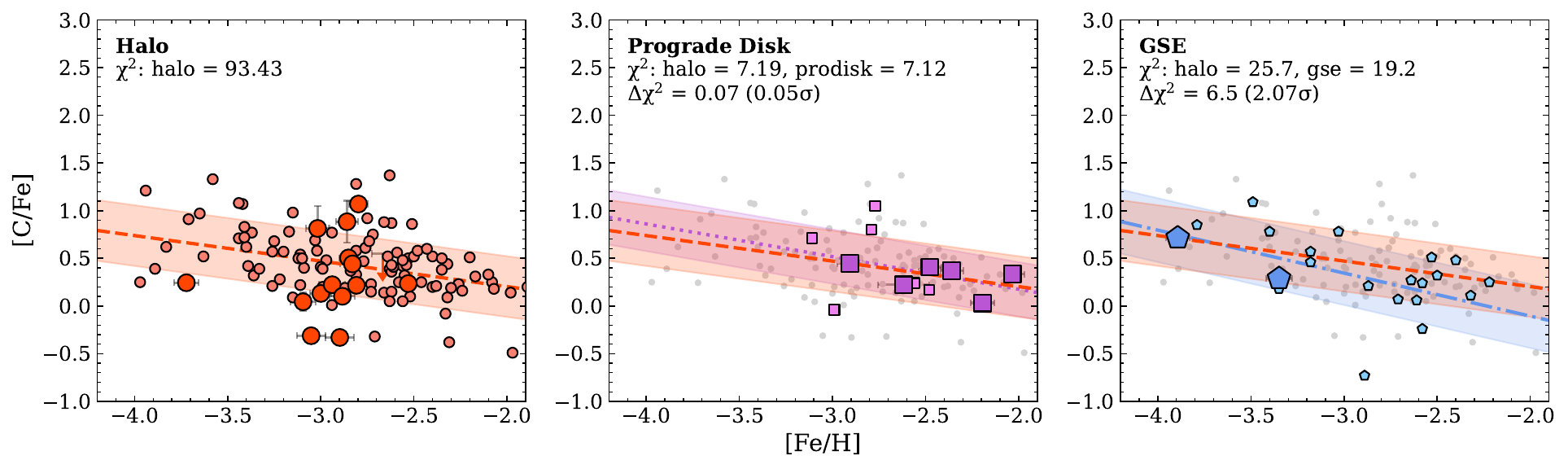}
    \end{subfigure}
    \hfill
    \begin{subfigure}{1\textwidth}
        \centering
        \includegraphics[width=1\linewidth]{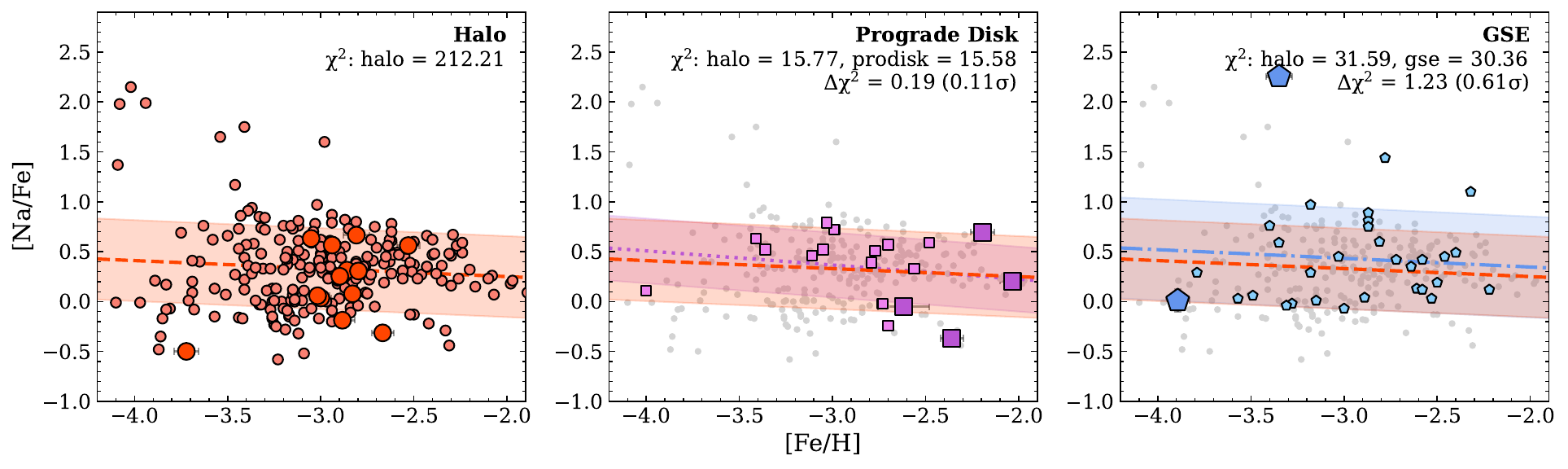}
    \end{subfigure}
    \hfill
    \begin{subfigure}{1\textwidth}
        \centering
        \includegraphics[width=1\linewidth]{images/deltachi2_plots/mg_abunds_no_retdisk_v3.pdf}
    \end{subfigure}
    \hfill
    \begin{subfigure}{1\textwidth}
        \centering
        \includegraphics[width=1\linewidth]{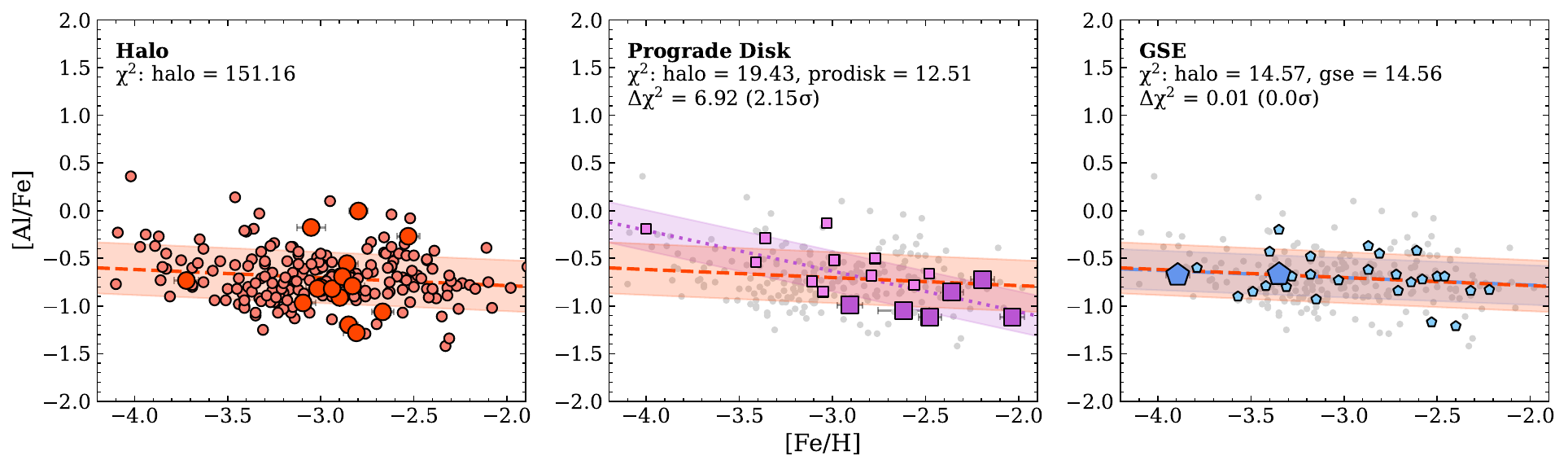}
    \end{subfigure}
    \caption{$\Delta \chi^2$ results for $\XFe{C}$ (first row), $\XFe{Na}$ (second row), $\XFe{Mg}$ (third row) and $\XFe{Al}$ (fourth row). The format is otherwise identical to Fig.~\ref{fig:scmg plot}.}
    \label{fig:delta chi2 plots 1}
\end{figure*}

\begin{figure*}  
    \centering
     \begin{subfigure}{\textwidth}
        \centering
        \includegraphics[width=1\linewidth]{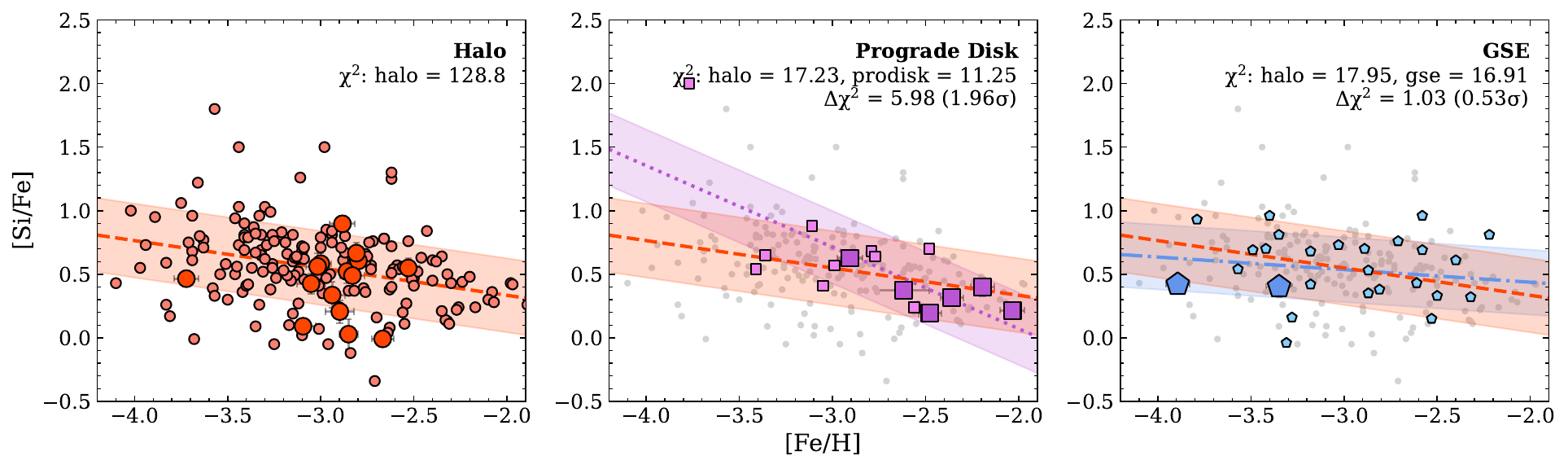}
    \end{subfigure}
    \hfill
    \begin{subfigure}{1\textwidth}
        \centering
        \includegraphics[width=1\linewidth]{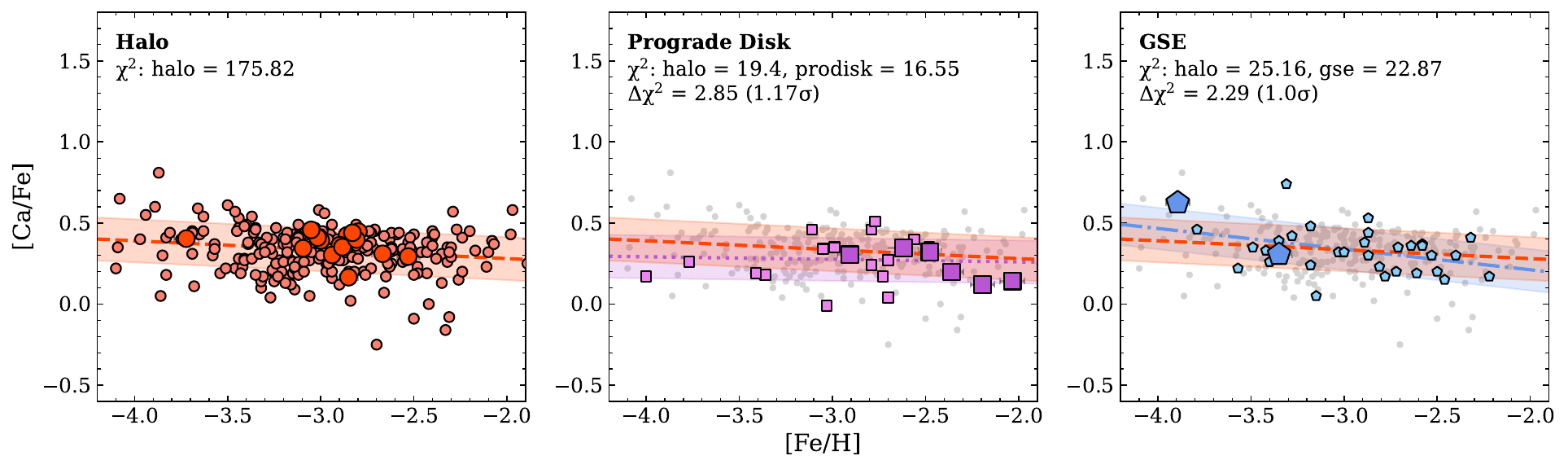}
    \end{subfigure}
    \hfill
    \begin{subfigure}{1\textwidth}
        \centering
        \includegraphics[width=1\linewidth]{images/deltachi2_plots/sc_abunds_no_retdisk_v3.pdf}
    \end{subfigure}
    \hfill
    \begin{subfigure}{1\textwidth}
        \centering
        \includegraphics[width=1\linewidth]{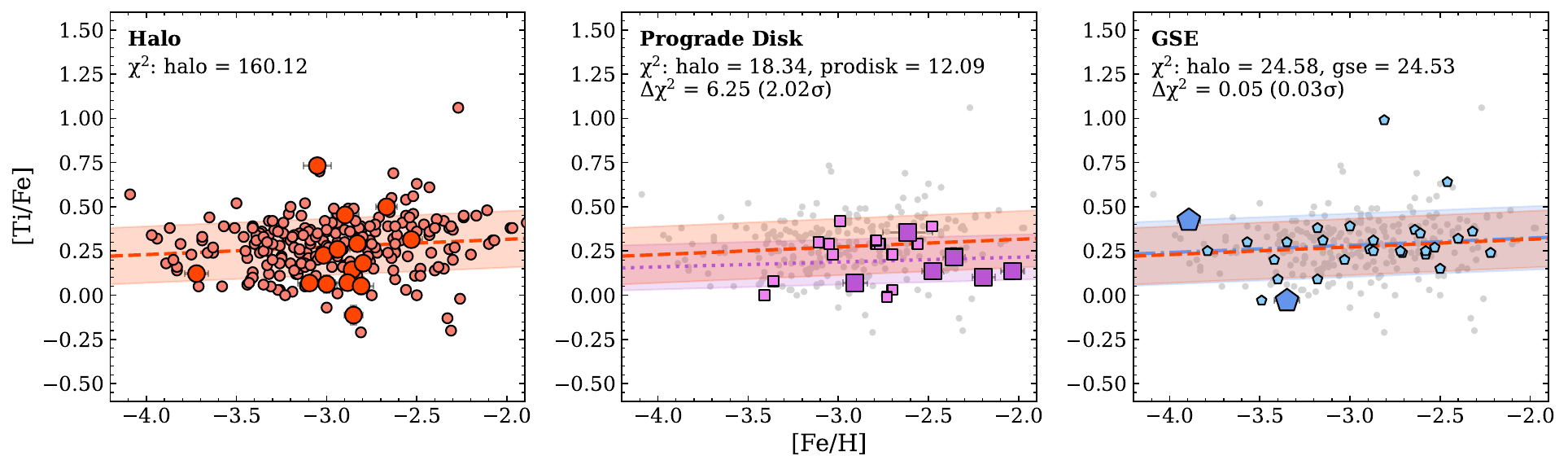}
    \end{subfigure}
    \caption{$\Delta \chi^2$ results for $\XFe{Si}$ (first row), $\XFe{Ca}$ (second row), $\XFe{Sc}$ (third row) and $\XFe{Ti}$ (fourth row). The format is otherwise identical to Fig.~\ref{fig:scmg plot}.}
    \label{fig:delta chi2 plots 2}
\end{figure*}

\begin{figure*}  
    \centering
     \begin{subfigure}{\textwidth}
        \centering
        \includegraphics[width=1\linewidth]{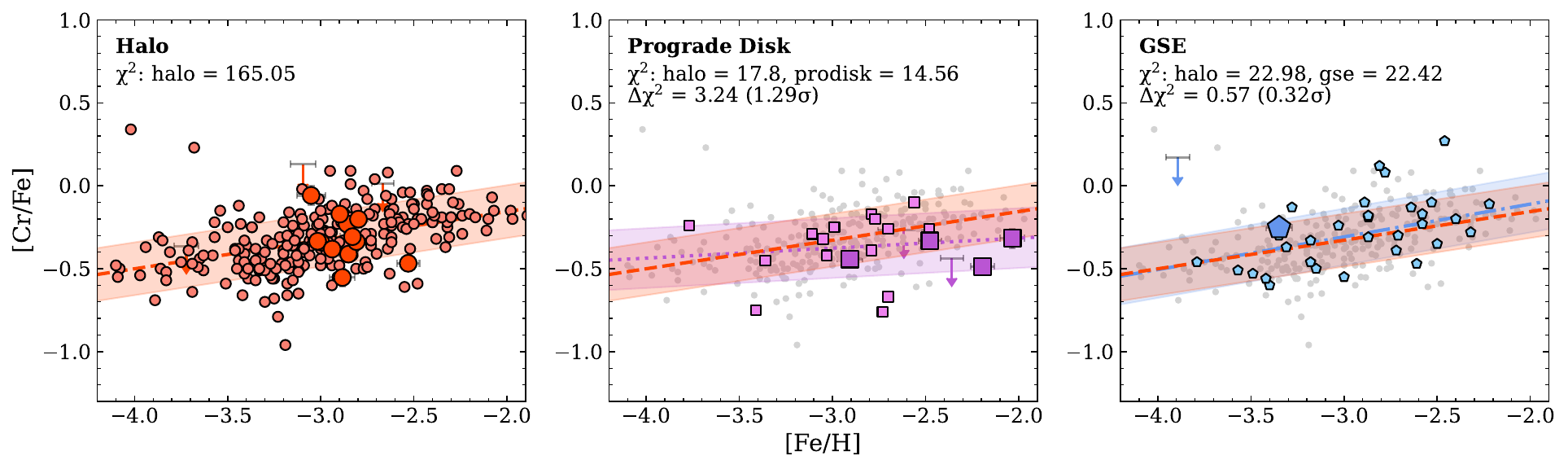}
    \end{subfigure}
    \hfill
    \begin{subfigure}{1\textwidth}
        \centering
        \includegraphics[width=1\linewidth]{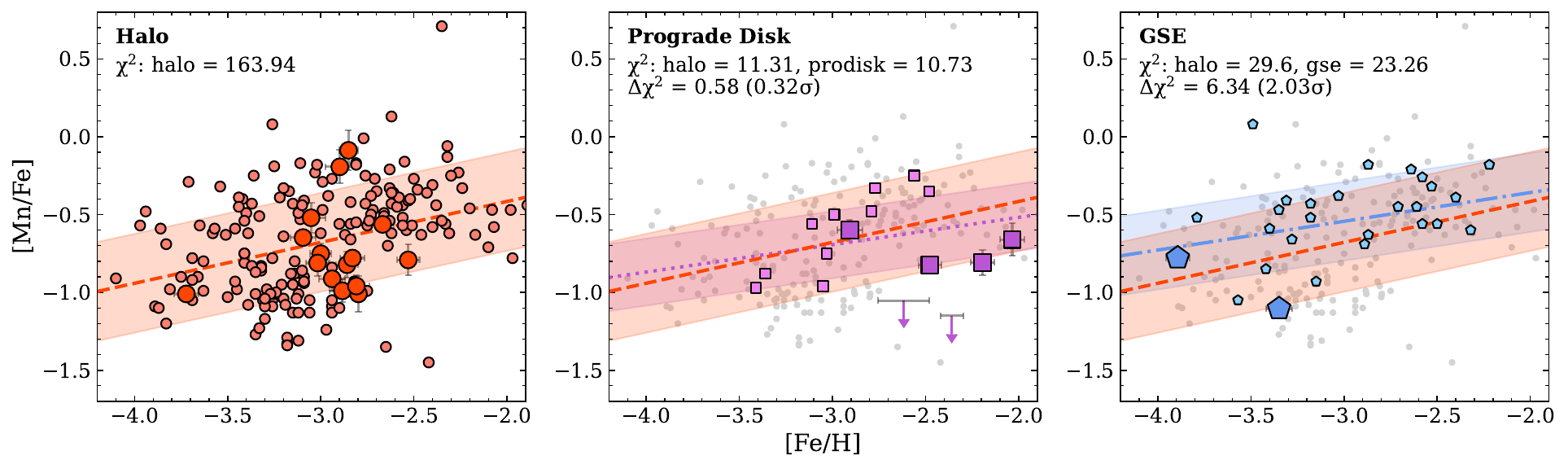}
    \end{subfigure}
    \hfill
    \begin{subfigure}{1\textwidth}
        \centering
        \includegraphics[width=1\linewidth]{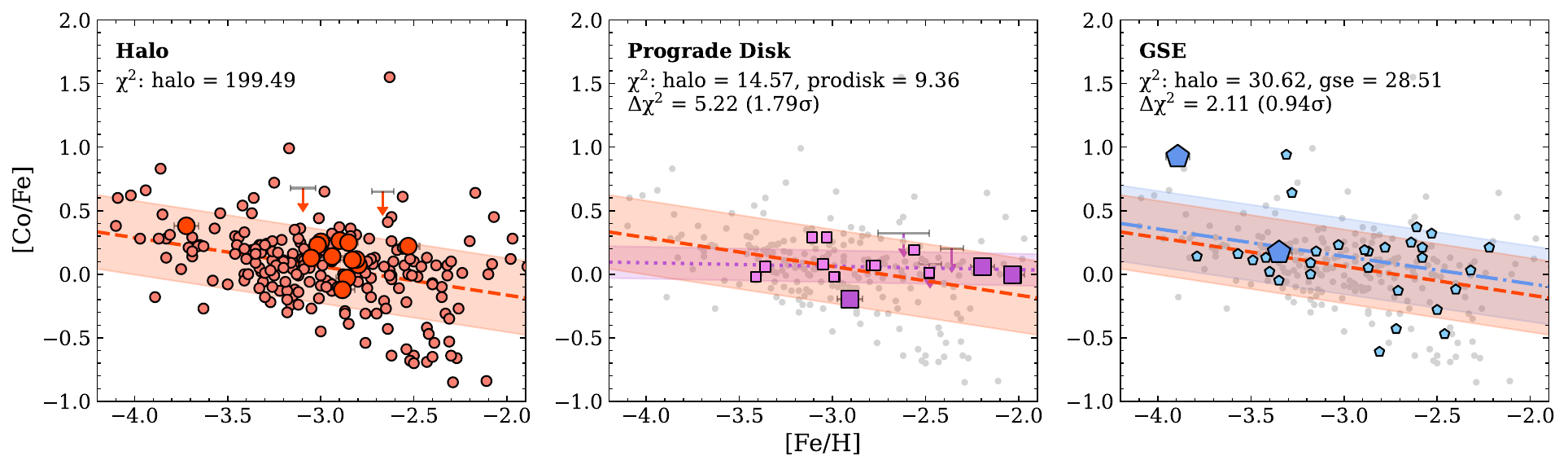}
    \end{subfigure}
    \hfill
    \begin{subfigure}{1\textwidth}
        \centering
        \includegraphics[width=1\linewidth]{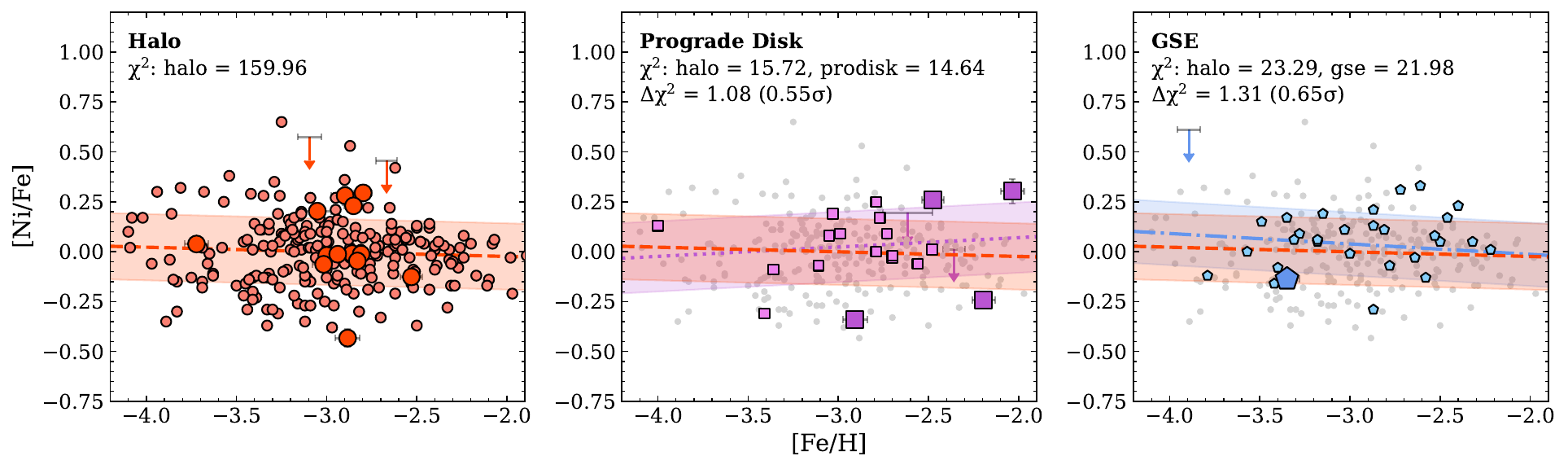}
    \end{subfigure}
    \caption{$\Delta \chi^2$ results for $\XFe{Cr}$ (first row), $\XFe{Mn}$ (second row), $\XFe{Co}$ (third row) and $\XFe{Ni}$ (fourth row). The format is otherwise identical to Fig.~\ref{fig:scmg plot}.}
    \label{fig:delta chi2 plots 3}
\end{figure*}

\begin{figure*}  
    \centering
     \begin{subfigure}{\textwidth}
        \centering
        \includegraphics[width=1\linewidth]{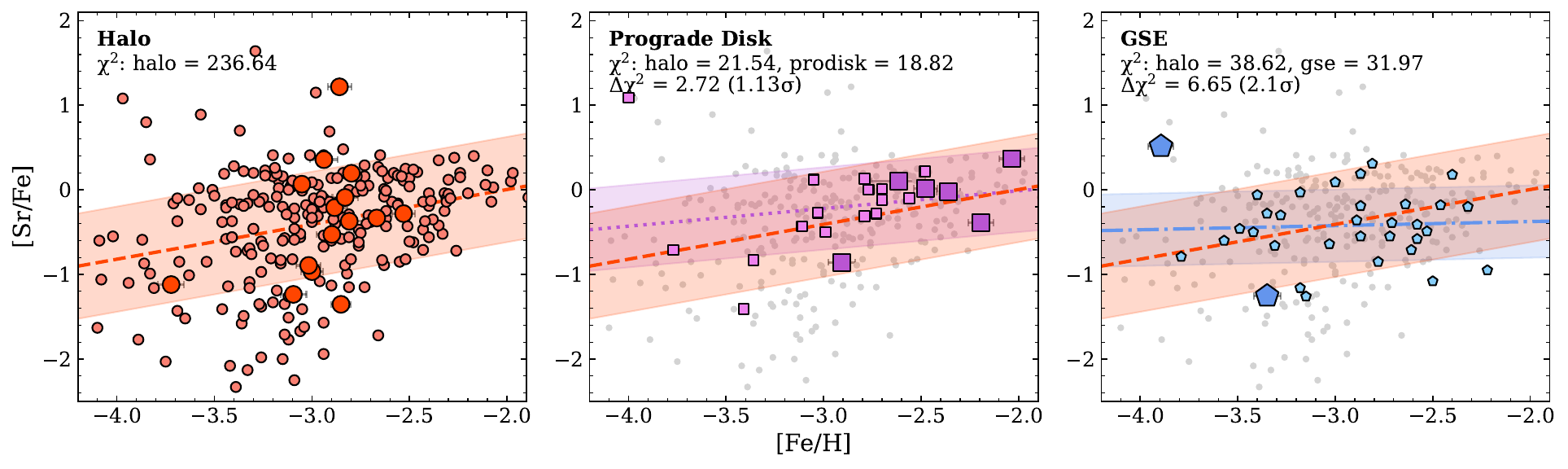}
    \end{subfigure}
    \hfill
    \begin{subfigure}{1\textwidth}
        \centering
        \includegraphics[width=1\linewidth]{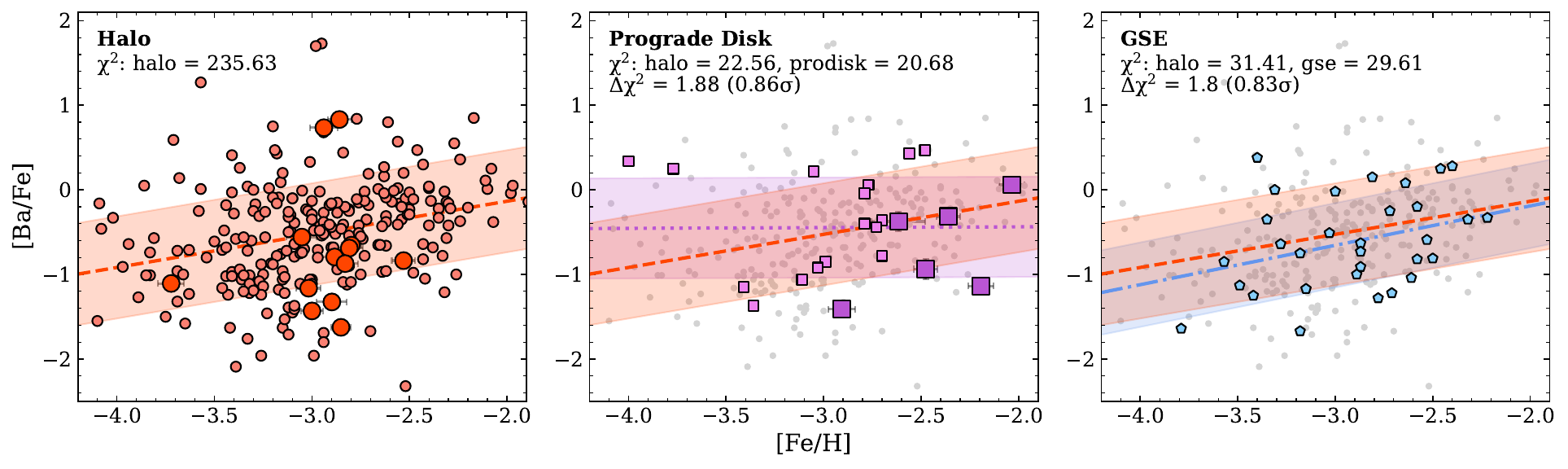}
    \end{subfigure}
    \hfill
    \begin{subfigure}{1\textwidth}
        \centering
        \includegraphics[width=1\linewidth]{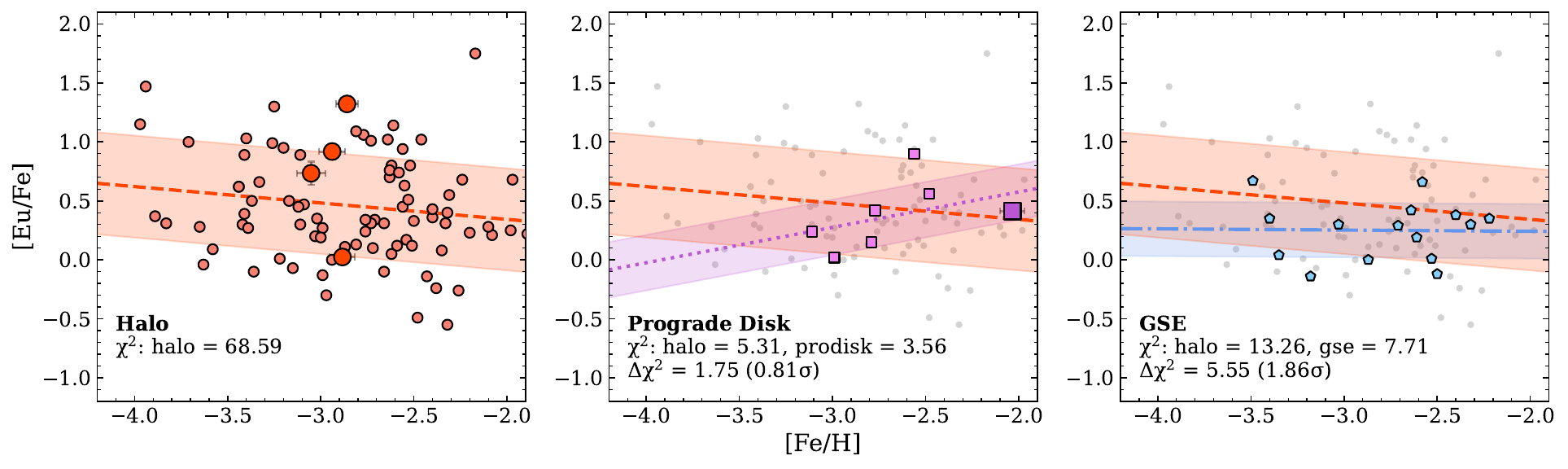}
    \end{subfigure}
    \caption{$\Delta \chi^2$ results for $\XFe{Sr}$ (first row), $\XFe{Ba}$ (second row) and $\XFe{Eu}$ (third row). The format is otherwise identical to Fig.~\ref{fig:scmg plot}.}
    \label{fig:delta chi2 plots 4}
\end{figure*}

Here, we show the $\Delta \chi^2$ results across our metallicities ($-4.2 \leq \FeH \leq -1.9$) for all our measured elements (excluding N). We show this for the prograde disk and GSE regions, with the retrograde disk excluded due to the small sample size. The halo is used as the null hypothesis against which to test the prograde disk/GSE regions for chemical diversity. In Fig.~\ref{fig:delta chi2 plots 1} we show the results for $\XFe{C}$, $\XFe{Na}$, $\XFe{Mg}$ and $\XFe{Al}$; in Fig.~\ref{fig:delta chi2 plots 2} we show the results for $\XFe{Si}$, $\XFe{Ca}$, $\XFe{Sc}$ and $\XFe{Ti}$; in Fig.~\ref{fig:delta chi2 plots 3} we show the results for $\XFe{Cr}$, $\XFe{Mn}$, $\XFe{Co}$ and $\XFe{Ni}$; and in Fig.~\ref{fig:delta chi2 plots 4} we show the results for $\XFe{Sr}$, $\XFe{Ba}$ and $\XFe{Eu}$. Plot format is identical to Fig.~\ref{fig:scmg plot}.

\bsp	
\label{lastpage}
\end{document}